\documentclass[12pt,a4paper]{article}
\usepackage{graphicx} 
\usepackage{physics}
\usepackage{jheppub}

\title{Bulk Reconstruction of Scalar Excitations in Flat$_3$/CCFT$_2$ and the Flat Limit from (A)dS$_3$/CFT$_2$}
\author[a,c]{Peng-Xiang Hao,}
\author[a]{Kotaro Shinmyo,}
\author[a]{Yu-ki Suzuki,}
\author[b]{Shunta Takahashi,}
\author[a,d]{Tadashi Takayanagi}

\affiliation[a]{
Center for Gravitational Physics and Quantum Information,
Yukawa Institute for Theoretical Physics, Kyoto University, Kyoto 606-8502, Japan}
\affiliation[b]{
Department of Physics, Kyoto University, Kyoto 606-8502, Japan
}
\affiliation[c]{Yau Mathematical Sciences Center, Tsinghua University, \\
Haidian District, Beijing 100084, China}
\affiliation[d]{Inamori Research Institute for Science,
620 Suiginya-cho, Shimogyo-ku, Kyoto 600-8411 Japan}
\date{}
\emailAdd{pxhao@yukawa.kyoto-u.ac.jp}
\emailAdd{kotaro.shinmyo@yukawa.kyoto-u.ac.jp}
\emailAdd{yu-ki.suzuki@yukawa.kyoto-u.ac.jp}
\emailAdd{shunta@gauge.scphys.kyoto-u.ac.jp}
\emailAdd{takayana@yukawa.kyoto-u.ac.jp}

\abstract{We explore the reconstruction of bulk local states in three-dimensional flat spacetime (Flat$_3$) using states from two-dimensional Carrollian conformal field theories (CCFT$_2$), proposed as dual field theories in one lower dimension. For massive scalar-type bulk excitations, reconstruction is achieved through states in the induced representation. This method successfully reproduces the bulk massive scalar spectrum and the bulk-to-bulk propagator. Additionally, we identify a new flat limit from AdS$_3$ and dS$_3$ spacetimes, further validating our proposal for scalar reconstruction in Flat$_3$/CCFT$_2$. }


\begin{document}
    \begin{flushright}
    YITP-25-83
    \\
    KUNS-3052
    \end{flushright}

    \maketitle

    \flushbottom

\newpage

\section{Introduction}
The flat space holography program aims to establish a holographic duality between gravity in asymptotically flat spacetime and a lower-dimensional field theory. The holographic principle \cite{tHooft:1993dmi,Susskind:1994vu} suggests that gravity within a specific spacetime region can be described on a lower-dimensional boundary. Extending this concept beyond the well-known AdS/CFT correspondence \cite{Maldacena:1997re,Gubser:1998bc,Witten:1998qj} to asymptotically flat spacetimes is one of the  significant areas of current research. Asymptotically flat spacetimes offer realistic models for a wide range of physical processes in our universe, from local phenomena to astrophysical scales. The asymptotic symmetries that preserve the boundary structure of these spacetimes form the BMS group \cite{Bondi:1962px,Sachs:1962wk}, an infinite-dimensional extension of the Poincaré group that includes supertranslations.

There are mainly two approaches to flat holography for arbitrary dimensions.\footnote{
Additionally, an alternative perspective on flat space holography, discussed in \cite{Ogawa:2022fhy}, is presented through wedge holography.} The first is the Flat/Carrollian CFT (CCFT, also known as BMS field theory) correspondence \cite{Barnich:2010eb,Bagchi:2010zz,Fareghbal:2013ifa}, where the field theory resides on the asymptotic boundary (null infinity), one dimension lower and including the time direction, following the typical codimension-one holographic duality. The second approach is celestial holography \cite{Pasterski:2016qvg,Pasterski:2017kqt}, where CFTs exist in a space two dimensions lower and without a time direction. Gravitational S-matrix elements, when expressed in a boost eigenstate basis, resemble conformal correlation functions. The relationship between these two approaches has been gradually evolving \cite{Donnay:2022wvx,Donnay:2022aba,Bagchi:2022emh,Bagchi:2023fbj}.

The main advantage of the celestial paradigm is its ability to apply various powerful CFT techniques easily. However, the Flat/CCFT approach introduces new challenges not commonly found in AdS/CFT correspondence. One challenge is the null boundary, which defines the field theory in a Newtonian geometry with one null direction, leading to many unusual Carrollian behaviors \cite{bacry1968possible,bergshoeff2014dynamics,Duval:2014uoa}. Another challenge is the radiative flux passing through the null boundary in higher-dimensional cases \cite{Trautman:1958zdi,Wald:1999wa}, leading
to the fact that the gravitational charges at null infinity are generically non-conserved, which indicates potentially additional sources in the dual field theory \cite{Donnay:2022aba}.

In spite of these difficulties, Flat/CCFT correspondence is developing rapidly in decades. The Bekenstein-Hawking entropy of flat cosmological horizons  matches with the Cardy formula in CCFTs \cite{Barnich:2012xq,Bagchi:2012xr}. The holographic entanglement entropy in flat holography is developed \cite{Li:2010dr,Bagchi:2014iea} , and the CCFT entanglement entropy calculated by twist operator or Rindler method can be compared to the swing surface holographic proposal \cite{Jiang:2017ecm}, which generalizes the RT proposal \cite{Ryu:2006bv,Ryu:2006ef} in AdS/CFT correspondence. The partition functions in flat holography are discussed in \cite{Barnich:2012rz,barnich2015one}, and the one-loop determinant is found to match the CCFT vacuum character. The connection between the Wald-Takahashi identities in CCFTs and the soft graviton theorem is discussed in \cite{Saha:2023hsl,Saha:2023abr}. Also, bulk massless particle provides a realization of the BMS operators \cite{Nguyen:2023vfz}, when pulled back to the conformal boundary, while the massive counterpart is explored in \cite{Have:2024dff}. Besides, as a special kind of quantum field theory, CCFT itself is rapidly advancing, with the developments on the correlation functions \cite{Bagchi:2009ca}, bootstrap procedure \cite{Bagchi:2016geg,Bagchi:2017cpu}, the universal properties on multiplets \cite{Chen:2020vvn,Chen:2022jhx}, the detailed models \cite{Hao:2021urq,Chen:2021xkw,Yu:2022bcp,Hao:2022xhq,Banerjee:2022ocj,deBoer:2023fnj}, and so on.

There are two widely used representations of the Carrollian conformal algebra in two dimensions. The highest weight representation of CCFT$_2$ \cite{Bagchi:2009ca} is similar to the Virasoro algebra, with $L_0$ bounded from below and yielding non-unitarity. The induced representation \cite{Barnich:2014kra,Barnich:2015uva}, seen as the ultra-relativistic limit of the highest weight representation, exhibits unitarity. Both representations play a role in the Flat/CCFT correspondence, though a clear dictionary is still lacking. One idea to advance this is to consider the bulk reconstruction of local excitations, known as bulk local states.

The reconstruction of  bulk local states in AdS/CFT correspondence has been widely studied \cite{Miyaji:2015fia,Nakayama:2015mva,Verlinde:2015qfa,Lewkowycz:2016ukf,Goto:2017olq}. It has been shown later \cite{Goto:2016wme} that this process is equivalent to the HKLL bulk reconstruction of operators \cite{Hamilton:2005ju,Hamilton:2006az}. Bulk local excitations can be expressed as linear combinations of CFT states in the highest weight module, with coefficients encoding the excitation's location in the AdS bulk. The bulk two-point function of the excitation can be reproduced by the inner product of the CFT state, and the AdS metric can be derived from the information metric of the CFT state by treating the bulk AdS coordinates as parameters in quantum estimation theory.

This narrative has recently been extended to the dS$_3$/CFT$_2$ correspondence \cite{Strominger:2001pn,Maldacena:2002vr,Hikida:2021ese} in \cite{Doi:2024nty}, where a straightforward analytical continuation fails due to the unusual conjugation relations of algebra generators in dS. This issue is resolved by a detailed analysis of state conjugation and the introduction of a CPT-invariant Euclidean vacuum state, which matches with the argument in \cite{Harlow:2023hjb} that the CPT should be gauged in de Sitter spaces. Consequently, the bulk two-point function and the dS metric can be derived from this CPT-invariant state, demonstrating the emergence of bulk time from the dual Euclidean CFT.

In the context of flat holography, the reconstruction of local excitations in the bulk is discussed in a recent study \cite{Chen:2023naw}. This work demonstrates that massless bulk states can be replicated in the highest weight states of CCFT, yielding the correct bulk two-point function. This approach is linked to the HKLL operator reconstruction in flat holography \cite{Donnay:2022wvx} due to the state-operator correspondence in the highest weight representation.

In this paper, we revisit the reconstruction of bulk local states in the flat holography. The proposal in \cite{Chen:2023naw} works well for massless cases but faces two main challenges when extended to massive cases. First, the correspondence between bulk massless states and CCFT states is not one-to-one, leaving conformal dimensions undetermined. Second, in the massive case, there are no solutions in the highest weight representation of the CCFT that satisfy the constraints on bulk local excitations.

To address these issues, we propose that the proper dual states to the bulk excitations are in the induced representation of the CCFT, rather than the highest weight representation. We demonstrate that the solution to the equations on bulk scalar excitations can be expressed as a linear combination of states in the induced representation. These states have a one-to-one correspondence with bulk massive scalars, allowing the quantum numbers to be fully determined. Additionally, the bulk two-point function and the flat metric can be reproduced from this solution in the induced representation.

We also consider the flat limit from AdS/CFT and dS/CFT to support our solution. By carefully examining the Killing vectors and algebra relations, we propose a new flat limit from (A)dS/CFT, which allows us to derive and confirm the solutions for the state and the bulk two-point function. This new flat limit may provide further insights into the flat holography.

The paper is structured as follows. In Section \ref{flat_ccft_exitation}, we provide a brief review of the Flat$_3$/CCFT$_2$ correspondence, emphasizing the two distinct representations in CCFT$_2$. We then set up our framework in global Minkowski spacetime, listing the Killing vectors and the equations for scalar excitations under a rotational invariance. This is followed by solutions in the induced representation, reproducing the bulk two-point function and flat metric. For completeness, we also discuss the solution in the highest weight representation for the massless case. In Section \ref{sec:flat_limit_AdS}, we analyze the flat limit from AdS$_3$/CFT$_2$ by examining the limit of Killing vectors in global AdS$_3$. We propose a new flat limit between CFT$_2$ generators and CCFT$_2$ generators, under which  the highest weight representation in the CFT becomes the induced representation in the CCFT. We then consider the behavior of the bulk local state and the bulk two-point function. In Section \ref{sec:flat_limit_dS}, we explore a similar flat limit from the dS$_3$/CFT$_2$ correspondence, finding consistent results for local excitations and the two-point function in flat holography. We conclude the paper in Section \ref{section: conclusion and outlook}, outlining future directions. Useful formulas are listed in Appendix \ref{append: formulas}, and technical details are provided in Appendix \ref{app: technical details}. In Appendix \ref{append: another solution}, we present an alternative solution for the bulk local scalar excitations in the induced representation, which  does not reproduce the bulk two-point function, for completeness.

\section{Scalar Excitation in Flat$_3$/CCFT$_2$ correspondence}\label{flat_ccft_exitation}

In this section, we will focus on the three-dimensional bulk and the reconstruction of local massive scalar excitations using states in the CCFT$_2$. Our approach involves examining the equations that the bulk excitations must satisfy and solving them within specific representations of the CCFT$_2$. A particular solution in the induced representation is chosen for the bulk reconstruction, as it aligns well with the bulk scalar spectrum and the  two-point function.

The structure of this section is as follows. In subsection \eqref{section: background}, we provide a concise review of the asymptotically flat spacetime in three-dimensional Einstein gravity and the two-dimensional Carrollian CFT, emphasizing the distinction between the induced representation and the highest weight representation, both of which are relevant to the CCFT$_2$ discussion. We then move to global Minkowski spacetime in subsection \eqref{section: global minkowski}, where we list its Killing vectors and formulate the equations that the scalar excitations must satisfy. These equations are solved in the induced representation in subsection \eqref{section: induced solution}, and we examine the properties of the solutions, including their ability to reproduce the bulk two-point function. Finally, we explore solutions in the highest weight representation in subsection \eqref{section: highest weight solution}, concluding that there are no suitable candidates for scalar reconstruction in this representation.
\subsection{Background of Flat$_3$/CCFT$_2$ correspondence}
\label{section: background}
The  algebra for the asymptotically flat spacetime in $d$-dimensions is the BMS$_d$ algebra. In the subsequent discussion, we will focus on the three-dimensional bulk and Einstein gravity, where the asymptotic algebra is expressed as
\begin{align}
    \begin{aligned}
        \left[L_{n},\,L_{m}\right] &= (n-m)L_{m+n}+\frac{c_L}{12} n(n^2-1)\delta_{m+n,0}, \\
	    \left[L_{n},\,M_{m}\right] &= (n-m)M_{m+n}+\frac{c_M}{12}  n(n^2-1)\delta_{m+n,0}, \\
	    \left[M_{n},\,M_{m}\right] &= 0.
    \end{aligned}
    \label{eq: bms_3 algebra}
\end{align}
The central charges read
\begin{equation}
c_L=0,\ \ c_M=\frac{3}{G},  \label{eq: BMS central charge}
\end{equation}
where $G$ denotes the gravitational constant in the three-dimensional spacetimes. The gravitational solutions can be expressed in the Bondi gauge,
\begin{equation}
ds^2=\Theta(\phi)du^2-2dudr+2\bigg(\Xi(\phi)+\frac{u\partial_\phi \Theta(\phi)}{2}\bigg)dud\phi+r^2d\phi^2,\ \ \phi\sim\phi+2\pi,
\end{equation}
where $u$ represents the retarded time, $r$ denotes the radial distance, and $\phi$ is the angle coordinate of the $S^1$. The constant mode solution is of particular interest due to its simplicity,
\begin{equation}\label{eq: const mode solution}
ds^2=Mdu^2-2dudr+2Jdud\phi+r^2d\phi^2,\ \ \phi\sim\phi+2\pi.
\end{equation}

These flat solutions can be obtained by taking the large AdS radius limit $l\rightarrow \infty$, or zero cosmological constant limit $\Lambda \rightarrow 0$ equivalently. This limit pushes the boundary cylinder to infinity as the length scale becomes infinite. Now the bulk looks like the previous center of the AdS. When $ M > 0$ , this represents a flat cosmological solution with a cosmological horizon at  $r_c = \frac{|J|}{\sqrt{M}}$. In the region $ r < r_c$ , closed timelike curves exist because  $\partial_\phi$  is timelike. The region $ r > r_c$ describes an expanding spacetime. When $ M < 0$ , the spinning particle creates a conical defect and a twist in the time identification. These configurations are locally flat but feature a delta source. The special case  $M = -1$  corresponds to the global Minkowski space. Other parameter values can be transformed back to the Minkowski metric, albeit with a different coordinate identification.
      \begin{equation}
      (u,\phi)\sim(u-2\pi r_0,\phi+2\pi\sqrt{-M}),\ \      r_0=\frac{J}{\sqrt{-M}}.
      \end{equation}

As $r\rightarrow\infty$, it reaches the null infinity,  which is a cylinder with the coordinates $(u,\phi)$.

From the perspective of the Flat/BMS correspondence (or Flat/CCFT correspondence), the proposed dual field theory is the BMS$_3$ field theory in two-dimensional spacetime, also known as BMSFT$_2$ or 2D Carrollian conformal field theory (CCFT$_2$). A CCFT$_2$ is a two-dimensional quantum field theory invariant under the following transformations,
\begin{equation}
\sigma\to f(\sigma), \quad \tau\to f^{\prime}(\sigma)\tau + g(\sigma).
\end{equation}

While the theory lacks Lorentz invariance, it still incorporates concepts of time and space. In CCFT, $\sigma$ represents a spatial direction, and $\tau$ signifies a time direction, albeit a null direction. This interpretation becomes more evident when considering CCFT$_2$ as the ultra-relativistic limit of a CFT$_2$ which is dual to AdS.

Now, let us consider a CCFT$_2$ on a cylinder parameterized by the coordinates $(\tau, \,\sigma)$ (corresponding to the coordinates $(u,\phi)$ on the boundary from the bulk viewpoint) with the identification
\begin{equation}
    \sigma\sim \sigma+2\pi.
\end{equation}
Then, the infinitesimal transformations are generated by the Fourier modes,
\begin{align}
L_{n} &= ie^{in\sigma }\partial_{\sigma} -n e^{in\sigma}\tau\partial_{\tau}\\
M_{n} &= ie^{in\sigma }\partial_{\tau}\label{lmncylinder}
\end{align}
which form the BMS$_3$ algebra \eqref{eq: bms_3 algebra} without the central terms. There are two quadratic Casimir operators,
\begin{equation}
C_1=M_0^2-M_{-1}M_1,\ \ C_2=2L_0M_0-\frac{1}{2}(L_{-1}M_1+L_1M_{-1}+M_1L_{-1}+M_{-1}L_1),   \label{eq: quadratic casimir in flat}
\end{equation}
which are important to study the algebra and the representations.

In the literature, two types of representations have been discussed: the highest weight representation \cite{Bagchi:2009ca} and the induced representation \cite{Barnich:2014kra,Barnich:2015uva}. The induced representation is unitary and can be derived from an ultra-relativistic limit of the highest weight representation of CFT$_2$s of the AdS dual. The highest weight representation of the BMS algebra is analogous to that of the Virasoro algebra, allowing techniques from CFT$_2$ to be adapted for BMSFT. However, the highest weight representation is non-unitary.

We begin by examining the induced representation. Given that the BMS algebra is the semi-direct sum of the Virasoro algebra and an Abelian ideal generated by $M_n$s, we can consider a representation derived from that of the ideal. Specifically, we focus on the special case where the states in an indecomposable representation meet the following conditions,
\begin{align}
    \begin{gathered}
        L_0|\Delta,\xi\rangle=\Delta|\Delta,\xi\rangle,\ \ M_0|\Delta,\xi\rangle=\xi|\Delta,\xi\rangle, \\
        M_n|\Delta,\xi\rangle=0,\ \ \forall n\neq0.
    \end{gathered}
    \label{eq: bms induced rep}
\end{align}
$\Delta$ and $\xi$ are referred to as the conformal weight and the boost charge of the state, respectively. The descendant states are generated by acting $L_{n\neq0}$ on it. This state has the following eigenvalues under the Casimir operators \eqref{eq: quadratic casimir in flat},
\begin{equation}\label{eq: induced casimir}
C_1|\Delta,\xi\rangle=\xi^2|\Delta,\xi\rangle,\ \ C_2|\Delta,\xi\rangle=2\Delta \xi|\Delta,\xi\rangle.
\end{equation}
Since the Casimir operators commute with all the generators by definition, the descendant states share the same eigenvalues. As was discussed in \cite{Barnich:2014kra, Barnich:2015uva, Campoleoni:2016vsh}, the induced representation is unitary.
Especially, the induced vacuum denoted as $|0_{\rm I}\rangle$ should satisfy
\begin{equation}\label{i72v}
L_0|0_{\rm I}\rangle=M_n|0_{\rm I}\rangle=0,\ \ \quad\quad \forall n\in \mathbb{Z}.
\end{equation}
In free models \cite{Hao:2021urq,Hao:2022xhq}, the induced vacuum acts as a direct product state, resulting in the ultra-local behavior of the correlation functions.

On the other hand, the highest weight representation of the BMS algebra is a direct generalization of the highest weight representation of the Virasoro algebra. By ensuring that $L_0$ is bounded below, the highest weight condition can be established,
\begin{align}
    \begin{gathered}
        L_0|\Delta,\xi\rangle=\Delta|\Delta,\xi\rangle,\ \ M_0|\Delta,\xi\rangle=\xi|\Delta,\xi\rangle, \\
        L_n|\Delta,\xi\rangle=M_n|\Delta,\xi\rangle=0,\ \ n>0,
    \end{gathered}
    \label{eq: bms highest weight rep}
\end{align}
which gives a highest weight primary state $|\Delta,\xi\rangle$.
The descendant states can be obtained by acting $L_{-n},M_{-n}$ with $n>0$ successively on it. The primary state together with its descendants forms a highest weight module, which shares the same eigenvalues of the Casimir operators,
\begin{equation}
C_1|\Delta,\xi\rangle=\xi^2|\Delta,\xi\rangle,\ \ C_2|\Delta,\xi\rangle=2(\Delta-1) \xi|\Delta,\xi\rangle
\end{equation}
Note that the induced representation and the highest weight representation have distinct eigenvalues for the Casimir operators mentioned above.

\subsection{Global Minkowski and equations for the scalar excitations}
\label{section: global minkowski}
\begin{figure}
    \centering
    \includegraphics[width=0.5\linewidth]{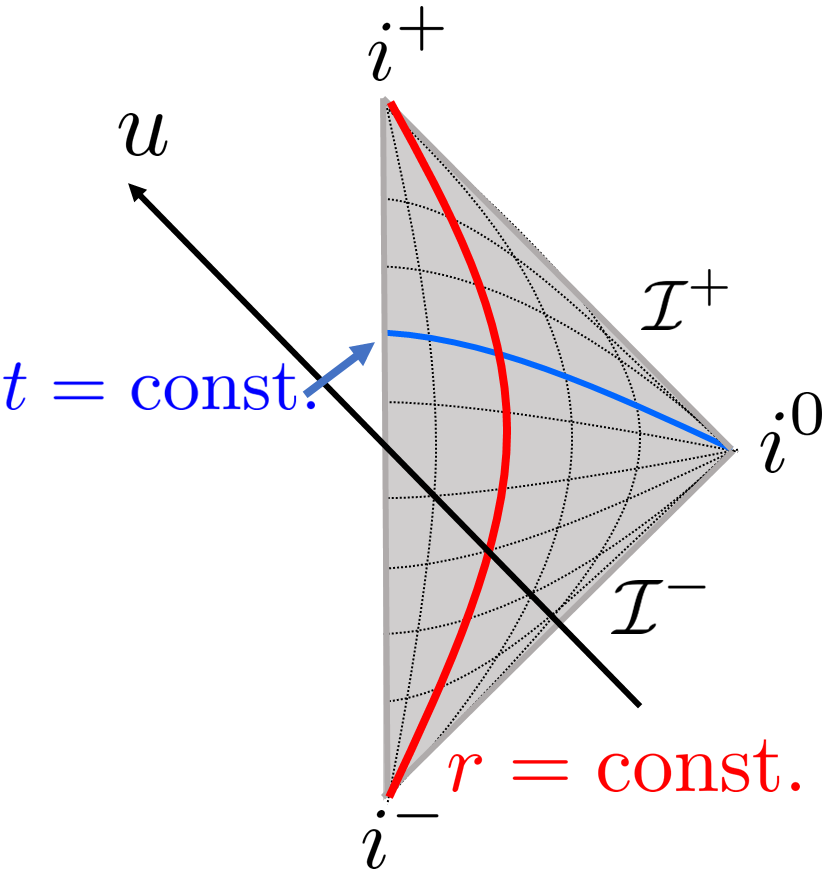}
    \caption{Penrose diagram of Mikowski space coordinated by \eqref{eq: 3d flat metric in the retarded coordinate}.}
    \label{fig:flat_coordinate}
\end{figure}

In the following discussion, we want to consider the bulk reconstruction of the scalar excitation in the global Minkowski spacetime in three dimensions. In the $(u,r,\phi) $ coordinate, it corresponds to the constant mode solution \eqref{eq: const mode solution} with $M=-1$,
\begin{equation}
ds^2=-du^2-2dudr+r^2d\phi^2,\ \ \phi\sim\phi+2\pi,\ \ 0\leq r<\infty,\ \ -\infty<u<\infty. \label{eq: 3d flat metric in the retarded coordinate}
\end{equation}
A further coordinate transformation
\begin{equation}\label{eq: global mink trans}
t=u+r,\ \ x=r\cos\phi,\ \ y=r\sin\phi,
\end{equation}
brings it back to the usual Cartesian coordinates
\begin{equation}
ds^2=-dt^2+dx^2+dy^2.
\end{equation}
Alternatively, in the polar coordinates, the flat metric reads,
\begin{equation}\label{eq: flat polar coordinate}
    ds^2=-dt^2+dr^2+r^2d\phi^2,\ \ \phi\sim\phi+2\pi,\ \ 0\leq r<\infty.
\end{equation}
The global Minkowski spacetime is maximally symmetric spacetime. In three dimensions, there are six Killing vectors, three of them are rotations and boosts,
\begin{equation}\label{eq: rotation and boosts}
    a_3=-y\partial_x+x\partial_y,\ \ a_2=y\partial_t+t\partial_y,\ \ a_1=x\partial_t+t\partial_x.
\end{equation}
The remaining three Killing vectors are translations,
\begin{equation}
    b_1=\partial_x,\ \ b_2=\partial_y,\ \ b_3=\partial_t.
\end{equation}
They form the Poincar\'{e} group $ISO(2,1)$, which can be also organized as $SL(2,\mathbb{R})\ltimes\mathbb{R}^3$, serving as the global part of the BMS$_3$ group, in the following combinations
\begin{equation}\nonumber
    L_1=ia_1- a_2,\ L_{-1}=ia_1+ a_2,\ L_0=i a_3,
\end{equation}
\begin{equation}
    M_1=b_1+ ib_2,\ M_{-1}=-b_1+ i b_2,\ M_0=i b_3.
\end{equation}
Then, the transformation \eqref{eq: global mink trans} gives the Killing vector in the $(u,r,\phi)$ coordinates, which will be used later,
\begin{equation}
    \begin{gathered}
        L_{\pm1}=-ie^{\pm i\phi}u\partial_u+ie^{\pm i\phi}(u+r)\partial_r\mp  e^{\pm i\phi}\Big(1+\frac{u}{r}\Big)\partial_\phi,\ \  L_0=i\partial_\phi, \\
        M_{\pm1}=\mp  e^{\pm i\phi}\partial_u\pm  e^{\pm i\phi}\partial_r+\frac{i}{r}e^{\pm i\phi}\partial_\phi,\ \ M_0=i\partial_u.
    \end{gathered}
   \label{eq: flat3killing}
\end{equation}
For the scalar functions $f(u,r,\phi)$ or $f(t,x,y)$ in the three-dimensional global Minkowski spacetime, we can consider the function space equipped with the $L^2$ inner product of the functions as usual
\begin{equation}
    \langle h,f\rangle=\int d^3x\sqrt{|g|}h^\ast f,
\end{equation}
where $\sqrt{|g|}$ is the volume form. Then, the Hermitian conjugation of the generators is expressed as follows
\begin{equation}    \label{eq: conjugation relation}
(L_n)^\dagger=L_{-n},\ \ (M_n)^\dagger=M_{-n}.
\end{equation}
These relationships can be verified through integration by parts in both the $(t,x,y)$ coordinate system and the $(u,r,\phi)$ coordinate system.

Additionally, it is crucial to consider the effect of the scalar function when acted upon by the Casimir operators,
\begin{equation}
C_1f(t,x,y)=(-\partial_t^2+\partial_x^2+\partial_y^2)f(t,x,y),\ \ C_2f(t,x,y)=0,
\end{equation}
\begin{equation}
C_1f(u,r,\phi)=\frac{\partial_\phi^2f}{r^2}+\frac{\partial_r f}{r}+\partial_r^2f-\frac{\partial_uf}{r}-2
   \partial_u\partial_rf,\ \ C_2f(u,r,\phi)=0,
\end{equation}
where the action under $C_2$ is always zero due to its scalar nature, while the action under $C_1$ can be compared to the Klein-Gordon equation for a massive scalar,
\begin{equation}\label{eq: kg equation}
    (C_1-m^2)f=0,\ \ C_2f=0.
\end{equation}
So the eigenvalue of $C_1$ should be associated with the parameter $m^2$, representing mass squared. In other words, the massive scalar excitations $|\phi(t,x,y)\rangle$ at the point $(t,x,y)$ in the bulk have specific eigenvalues under the Casimir operator, suggesting that we should reconstruct them in the boundary theory using states from a single module, whose quantum numbers are also labeled by these eigenvalues. We will discuss these relations separately in both the induced representation and the highest weight representation in the following subsections.

Before delving into detailed representations, we will list the equations that a scalar excitation must satisfy due to rotational invariance. Consider a bulk local excitation at the origin $(t,x,y)=(0,0,0)$, which is the fixed point of the rotations and boosts \eqref{eq: rotation and boosts}. Therefore, the dual CCFT$_2$ scalar excitation $|\phi(0,0,0)\rangle$  must adhere to the following constraints imposed by the BMS generators,
\begin{equation}\label{eq: origin equation}
L_{0,\pm1}|\phi(0,0,0)\rangle=0.
\end{equation}
In general, for a point $(t,0,0)$, the rotational invariance conditions give the equations for a general scalar type excitation $|\phi(t,0,0)\rangle$,
\begin{equation}\label{eq: t equation}
(L_1-itM_1)|\phi(t,0,0)\rangle=(L_{-1}+itM_{-1})|\phi(t,0,0)\rangle=L_0|\phi(t,0,0)\rangle=0
\end{equation}
These conditions can be verified by examining the vanishing action of the linear combination of global BMS generators at the point $(t,0,0)$. Additionally, these conditions align with the algebra in the following way. The excitation at any point can be linked to the excitation at the origin through translation symmetries,
\begin{equation}\label{eq: generic location}
|\phi(t,x,y)\rangle=e^{iM_0t}e^{-iy/2(M_1+M_{-1})}e^{x/2(M_1-M_{-1})}|\phi(0,0,0)\rangle.
\end{equation}
We can then use algebra to move the conditions from the origin to the point $(t,0,0)$. Applying the BCH formula, we get
\begin{equation}
    e^{-aM_0}e^{bL_1}e^{aM_0}=e^{bL_1+abM_1}.
\end{equation}
Differentiating with respect to\ $b$ and setting $b=0$, we have
\begin{equation}
    e^{-aM_0}L_1e^{aM_0}=L_1+aM_1.
\end{equation}
With the above expression, we can check that the condition in \eqref{eq: origin equation}
\begin{equation}
   (L_1-itM_1)|\phi(t,0,0)\rangle=0
\end{equation}
is consistent with the condition in \eqref{eq: t equation}
\begin{equation}
    L_1|\phi(0,0,0)\rangle=0
\end{equation}
and the relation \eqref{eq: generic location} with $x=y=0$
\begin{equation}
    |\phi(0,0,0)\rangle=e^{-itM_0}|\phi(t,0,0)\rangle.
\end{equation}
A parallel discussion on the remaining conditions is similar and will not be repeated here.

\subsection{Solution in the induced representation}
\label{section: induced solution}
We have outlined the equations for the massive scalar excitations $|\phi\rangle$ in the bulk. Next, we aim to represent this state as a linear combination of states in the boundary CCFT$_2$, and determine the coefficients by solving the conditions given in equations \eqref{eq: origin equation} and \eqref{eq: t equation}.

In this subsection, we will focus on the solution with sensible properties that align well with the bulk results. For completeness, an alternative solution using induced representation, which is not suitable for bulk reconstruction, is provided in appendix \ref{append: another solution}.

We begin with the states in the induced representation in CCFT$_2$. By comparing the eigenvalues of the Casimir for the bulk scalar field \eqref{eq: kg equation} and those of the induced module at the boundary \eqref{eq: induced casimir}, we establish the following relations,
\begin{equation}
    m^2=\xi^2,\ \ 2\Delta\xi=0.
\end{equation}
If all possible mass parameters are considered, from the above relations we have the dictionary as
\begin{equation}
    \xi=\pm m,\ \ \Delta=0, \label{eq: xi and weight for scalar in bms}
\end{equation}
which relates the weight and the boost charge of the induced module in the CCFT$_2$ to the mass parameter of the scalar excitation in the bulk. Then, we have the ansatz for the bulk state
\begin{equation}\label{eq: excitation ansatz induced}
    |\phi\rangle=\sum_{i,j} c_{ij} L_{-1}^iL_{1}^j|\xi\rangle,\ \
\end{equation}
where $|\xi\rangle=|\Delta=0,\xi\rangle$ is the induced primary state and $L_{-1}^iL_{1}^j|\xi\rangle$ form a basis in the module. Currently, we have not specified the position of the bulk excitation. When we start with the excitation at a certain point, we observe that it still satisfies the ansatz after moving to another point, as described by the relation \eqref{eq: generic location}.

Next, we examine the constraints on the excitation  \eqref{eq: t equation} at $(t,0,0)$. Using the ansatz \eqref{eq: excitation ansatz induced}, these conditions transform into
\begin{equation}\label{eq: t equation with ansatz}
    \sum_{i,j} c_{ij}(L_1-itM_1) L_{-1}^iL_{1}^j|\xi\rangle=  \sum_{i,j} c_{ij}(L_{-1}+itM_{-1}) L_{-1}^iL_{1}^j|\xi\rangle=\sum_{i,j} c_{ij}L_{0} L_{-1}^iL_{1}^j|\xi\rangle=0.
\end{equation}
The last equation indicates that $i=j$ which reduces $c_{ij}$ to $c_i\delta_{i,j}$. Then we have
\begin{equation}
    |\phi(t,0,0)\rangle=\sum_{k=0}^\infty c_k|k\rangle, \ \ |k\rangle\equiv L_{-1}^kL_1^k|\xi\rangle.
\end{equation}
With the help of the formulae \eqref{eq: action of generators} in Appendix \ref{append: formulas} derived from the BMS algebra \eqref{eq: bms_3 algebra}, we find the solution to the given equations \eqref{eq: t equation with ansatz}
\begin{equation}\label{eq: t solution}
    |\phi(t,0,0)\rangle=e^{i \xi
   (t-1)}\sum_k\frac{2^{-k} \xi ^{-k} \left(-\frac{i}{t}\right)^{k+1} }{k!}|k\rangle
\end{equation}
This is the key result of this section. A special overall constant is chosen for later convenience. As a self-consistent check, it satisfies
\begin{equation}\label{eq: t00 in 100}
    |\phi(t,0,0)\rangle=e^{iM_0(t-1)}|\phi(1,0,0)\rangle.
\end{equation}
Considering the excitation at $(t,0,0)$ instead of the origin offers two key advantages, as shown in the solution \eqref{eq: t solution}. Firstly, it provides a more general solution for arbitrary time $t$. Secondly, our ansatz \eqref{eq: excitation ansatz induced} implicitly assumes that the coefficient should be non-singular. Note that our solution \eqref{eq: t solution} becomes singular as $t \to 0$, making it impossible to solve the equation \eqref{eq: origin equation} to obtain this solution. Then, one can relate the general scalar excitation at arbitrary point to the solution \eqref{eq: t solution},
\begin{equation}\label{eq: general solution}
|\phi(t,x,y)\rangle=e^{-iy/2(M_1+M_{-1})}e^{x/2(M_1-M_{-1})}|\phi(t,0,0)\rangle.
\end{equation}

With the bulk reconstruction of local scalar excitations given by equation \eqref{eq: general solution}, we aim to delve deeper into its aspects, including the conjugate state, the bulk two-point function, and the information metric.
\paragraph{Conjugate state} We aim to observe the behavior of the excitation \eqref{eq: general solution} under conjugation, specifically considering the Hermitian conjugation relation of the generators \eqref{eq: conjugation relation}. The conjugate state to the basis $|k\rangle$ becomes
\begin{equation}
    \langle k|=(|k\rangle)^\dagger=\langle \xi|L_{-1}^kL_{1}^k
\end{equation}
where $\langle \xi|$ is out state conjugate to the induced primary state $|\xi\rangle$. It is also an induced primary state, satisfying
\begin{equation}
    \langle \xi|L_0=0,\ \ \langle \xi|M_0=\xi,\ \ \langle \xi|M_n=0,\ \ \forall n\neq 0.
\end{equation}
Then, the conjugate state at $(t,0,0)$ is
\begin{equation}
    \langle \phi(t,0,0)|=(|\phi(t,0,0)\rangle)^\dagger=\langle k|e^{-i \xi
   (t-1)}\sum_k\frac{2^{-k} \xi ^{-k} \left(\frac{i}{t}\right)^{k+1} }{k!},
\end{equation}
which is the solution to
\begin{equation}
\langle \phi(t,0,0)|(L_1-itM_1)=\langle \phi(t,0,0)|(L_{-1}+itM_{-1})=\langle \phi(t,0,0)|L_0=0.
\end{equation}
Ultimately, we determine the conjugate state to the general excitation at any given point,
\begin{equation}
    \langle \phi(t,x,y)|=\langle\phi(t,0,0)|e^{-x/2(M_1-M_{-1})}e^{iy/2(M_1+M_{-1})}.
\end{equation}
\paragraph{Inner product and dual basis}

Before delving into the bulk two-point functions, let's take a moment to examine the inner product in the induced representation in CCFT$_2$, which is quite subtle and differs from the usual case. Having an induced primary state $|\xi\rangle$ and its conjugate state $\langle\xi|$, we want to have some idea on the inner product in this module. The descendant bra and ket states are
\begin{equation}\label{eq: ij state}
    \langle i,j|=\langle \xi|L_{-1}^jL_1^i,\ \ |i,j\rangle=L_{-1}^iL_1^j|\xi\rangle.
\end{equation}
A general inner product
\begin{equation}
    N_{ijkl}=\langle i,j|k,l\rangle=\langle \xi|L_{-1}^jL_1^iL_{-1}^kL_1^l|\xi\rangle
\end{equation}
is hard to compute since all the $L_n$ are creation operators, which is significantly different from the usual case in the highest weight representation in CFT$_2$. To overcome this difficulty, it is useful to introduce another basis to expand the bra states, which is called the dual basis, defined as
\begin{equation}\label{eq: dual basis1}
    \ ^\vee\!\langle i,j|=(N^{-1})_{i'j'kl}\langle i',j'|,
\end{equation}
so that it is orthogonal
\begin{equation}\label{eq: dual basis2}
    \ ^\vee\!\langle i,j|k,l\rangle=\delta_{i,k}\delta_{j,l}.
\end{equation}
Note that $(N^{-1})_{i'j'kl}$ represents an element of the inverse matrix of the inner product. If $N$ is known, the relationship between the dual basis and the original $\langle i,j|$ basis can be explicitly calculated. However, since $N$ is unknown here, the definition in \eqref{eq: dual basis1} is merely formal. Therefore, \eqref{eq: dual basis2} should be considered as the guiding principle for finding the dual basis.

In practice, chosen the basis of the ket states as $|k\rangle$, the dual basis $\ ^\vee\!\langle k|$ is defined as follows. The 'primary state' in the dual basis $\ ^\vee\!\langle 0|$  is
\begin{equation}
    \ ^\vee\!\langle 0|L_{0}=\ ^\vee\!\langle 0|\Delta=0,\ \  \ ^\vee\!\langle 0|L_{\pm1}=0,
\end{equation}
which belongs to the same module of $\langle \xi|$ and the ``descendant states" are 
            \begin{equation}\label{eq: dual descendant states}
                \ ^\vee\!\langle k|=\sum_{r=k}^{\infty}(-1)^r(2\xi)^{2k-2r}C_{kr}\ ^\vee\!\langle 0|M_1^rM_{-1}^r,\qquad \xi\neq0,
            \end{equation}
            where
            \begin{equation}\label{eq: Ckr coefficient}
                C_{0r}:=\delta_{r,0},\qquad C_{kr}:=\frac{k}{2r-k}\binom{2r-k}{r-k}\quad(r\geq k\geq 1).
            \end{equation}
After some algebra (See Appendix \ref{subappend:orthogonality}), it follows
\begin{equation}\label{eq: inner product}
    \ ^\vee\!\langle k'|k\rangle=N_k\delta_{k,k'},\quad
    N_k=(k!(2\xi)^k)^2,
\end{equation}
which checks the definition \eqref{eq: dual basis2}. Note that the dual basis is the re-organization of the usual basis of the bra states with the usual conjugate relation. We give a toy model to illustrate the dual basis in appendix.\,\ref{app: toy model}.
\paragraph{Two-point function} In the bulk, the bulk-to-bulk propagator is the two-point function $G(t,x,y)=\langle O(t,x,y)O(0,0,0)\rangle$ of the massive scalar $O$, satisfying
\begin{equation}
(C_1-m^2)G(t,x,y)=0,
\end{equation}
which has the solution
\begin{equation}
G(t,x,y)=-\frac{e^{\pm m D_{\text{flat}}}}{4\pi D_{\text{flat}}},
\end{equation}
where $D_{\text{flat}}=\sqrt{-t^2+x^2+y^2}$ is the geodesic distance between the two points $(t,x,y)$ and $(0,0,0)$ in the flat$_3$ spacetime. For other Green's functions like the retarted or Feynman one, we can construct properly from this Wightman function. For sufficiently large spacelike separations, the two-point function should approach zero. Consequently, the solution with $-m$ should be selected as
\begin{equation}\label{eq: bulk 2pt}
G(t,x,y)=-\frac{e^{- m D_{\text{flat}}}}{4\pi D_{\text{flat}}}.
\end{equation}
It is anticipated that the two-point function will be accurately reproduced by our reconstruction of the bulk scalar \eqref{eq: t solution},
\begin{equation}
    G(t,x,y)\overset{?}{=}\langle \phi(0,0,0)|\phi(t,x,y)\rangle,
\end{equation}
where $\langle \phi(0,0,0)|$ is the conjugate state at the origin. There are two main challenges to address. First, the scalar excitation state at the origin has a divergent coefficient, which can be resolved through the translational symmetries of the two-point function. However, the second issue is more critical. The states are represented in the basis as descendant states in the induced module, yet their inner product cannot be calculated as previously discussed. Fortunately, we have the dual state $\ ^\vee\!\langle0|$, satisfying
\begin{equation}
    ^\vee\langle0|L_{0,\pm1}=0,
\end{equation}
which coincident with the condition for the scalar excitation at origin, conjugate to \eqref{eq: origin equation}. A natural question is whether this dual state $\ ^\vee\!\langle0|$ serves as the conjugate state of the scalar excitation at the origin,
\begin{equation}
    \langle\phi(0,0,0)|\overset{?}{\sim}\ ^\vee\!\langle0|.
\end{equation}
We will verify this by calculating the inner product between $\ ^\vee\!\langle0|$ and $|\phi(t,x,y)\rangle$ explicitly. First, we examine the time separation case, where the inner product is $\ ^\vee\!\langle0|\phi(t,0,0)\rangle$. To proceed, we denote
\begin{equation}
    f_{k',k}(a)=\ ^\vee\!\langle k '|e^{aM_0}|k\rangle.
\end{equation}
Its derivative with respect to $a$ is
\begin{equation}
    \partial_a f_{k',k}(a)=\ ^\vee\!\langle k '|e^{aM_0}M_0|k\rangle=-2k^2\xi \ ^\vee\!\langle k '|e^{aM_0}|k-1\rangle+\xi\ ^\vee\!\langle k '|e^{aM_0}|k\rangle,
\end{equation}
where we have used \eqref{eq: m0 action on k} in the last equation. This in turn gives the recursive relation of $f_{k',k}(a)$,
\begin{equation}
    \partial_a f_{k',k}(a)=-2k^2\xi f_{k',k-1}(a)+\xi f_{k',k}(a).
\end{equation}
With the initial condition
\begin{equation}
    f_{0,0}(a)=e^{a\xi},
\end{equation}
we find that the solution to the recursive relation is
\begin{equation}
f_{k',k}(a)=\langle k'|e^{aM_0}|k\rangle=\frac{(k'+1)_{k-k'}^2}{\Gamma(k-k'+1)}(-2a\xi)^{k-k'}e^{a\xi}N_{k'},
\end{equation}
where $N_{k'}$ is the normalization constant in eq.\,\eqref{eq: inner product}. Especially, we have
\begin{equation}
f_{0,k}(a)=\langle 0|e^{aM_0}|k\rangle=\Gamma(k+1)(-2a\xi)^ke^{a\xi}.
\end{equation}
Then, the inner product is
\begin{equation}
    \ ^\vee\!\langle0|\phi(t,0,0)\rangle= \ ^\vee\!\langle0|e^{iM_0(t-1)}|\phi(1,0,0)\rangle=\sum_k\frac{2^{-k} \xi ^{-k} \left(-i\right)^{k+1} }{k!}f_{0,k}(i(t-1)),
\end{equation}
where we used the relation \eqref{eq: t00 in 100} and the explicit expression for $|\phi(1,0,0)\rangle$,
\begin{equation}
    |\phi(1,0,0)\rangle=\sum_k\frac{2^{-k} \xi ^{-k} \left(-i\right)^{k+1} }{k!}|k\rangle.
\end{equation}
After the summation, we have
\begin{equation}
    \ ^\vee\!\langle0|\phi(t,0,0)\rangle=-\frac{ie^{i(t-1)\xi}}{t}.
\end{equation}
This is exactly the $k=0$ coefficient in the $|\phi(t,0,0)\rangle$ solution \eqref{eq: t solution}. Indeed, this inner product can be calculated by the solution \eqref{eq: t solution} and the orthogonal relation \eqref{eq: inner product}. The introduction of the calculation from $f_{k'.k}(a)$ and $|\phi(1,0,0)\rangle$ serves three purposes. Firstly, it acts as a consistency check. Secondly, it clarifies the overall coefficient in the solution \eqref{eq: t solution}. Lastly, it provides the action of $e^{aM_0}$ on the $|k\rangle$ state \eqref{eq: eam0 action on k}, which will be utilized in subsequent calculations. Then, we can propose\footnote{Note added (2026/03/14): The introduction of the dual basis may appear somewhat artificial at this stage. However, after this work was completed, it received a theoretical justification via a more detailed flat limit analysis presented in Ref. \cite[Section 2]{Hao:2026cqm}.}
\begin{equation}\label{eq: relation between conjugate state}
\langle \phi(0,0,0)|=-\frac{e^{i\xi}}{4\pi}\ ^\vee\!\langle 0|,
\end{equation}
so that the two-point function reads
\begin{equation}
    G(t,0,0)=-\frac{e^{i\xi t}}{4\pi t}.
\end{equation}
We then consider a more general case where the excitation is located at $(t,x,0)$,
\begin{equation}
|\phi(t,x,0)\rangle=e^{x/2(M_1-M_{-1})}|\phi(t,0,0)\rangle.
\end{equation}
To proceed, we denote
\begin{equation}
\ ^\vee\!\langle0|e^{a(M_1-M_{-1})}|i,j\rangle=g_{i,j}(a),
\end{equation}
where the state $|i,j\rangle$ is defined in \eqref{eq: ij state}. Similar to the discussion on $f_{k',k}(a)$, with the formula \eqref{eq: action of generators}, we have the recursive relation on $g_{i,j}(a)$,
\begin{equation}
\partial_a g_{i,j}(a)=-2\xi i(i-1)jg_{i-2,j-1}(a)+2\xi ig_{i-1,j}(a)+2\xi j g_{i,j-1}(a).
\end{equation}
With the initial condition
\begin{equation}
g_{0,0}=1,\ \ g_{n,m}=0,\ \ g_{m,n}=0,\ \ \forall n<0,
\end{equation}
We have the solution
\begin{equation}
g_{k,k}(a)=\sum_{i=0}^{k}\frac{(k-2i+1)_{2i}(k-i+1)_{i}}{\Gamma(i+1)}(2a\xi)^{2(k-i)}(-1)^i.
\end{equation}
Thus, we obtain
\begin{equation}
    \ ^\vee\!\langle0|\phi(t,x,0)\rangle=\ ^\vee\!\langle0|e^{x/2(M_1-M_{-1})}|\phi(t,0,0)\rangle=\sum_{k=0}^{\infty}g_{k,k}\left(\frac{x}{2}\right)c_k,
\end{equation}
where $c_k$ is the coefficients in \eqref{eq: t solution}. This is a double summation which is very hard to solve. We first consider the massless case $\xi=m=0$, where only one term for each even $k$ survives. The summation can be done
\begin{equation}
\langle \phi(0,0,0)|e^{\frac{x}{2}(M_1-M_{-1})}|\phi(t,0,0)\rangle=-\frac{1}{4\pi\sqrt{x^2-t^2}}.
\end{equation}
Inspired by this procedure, we can proceed by grouping the terms with different powers of $\xi$ and reorganizing the summation in $g_{k,k}(a)$ with $i$ replaced by $(k-j)/2$,
\begin{equation}
g_{k,k}=\sum_{j}\frac{i^{k-j} 2^{j+k} \Gamma (k+1)^2 (a \xi )^{j+k}}{\Gamma (j+1) \Gamma
   \left(\frac{1}{2} (-j+k+2)\right) \Gamma \left(\frac{1}{2} (j+k+2)\right)}.
\end{equation}
Then, we can exchange the order of the summation and we have
\begin{equation}
\langle \phi(0,0,0)|e^{\frac{x}{2}(M_1-M_{-1})}|\phi(t,0,0)\rangle=\frac{e^{i\xi}}{4\pi}\sum_{j}-\frac{i \xi ^j \left(i \left(\sqrt{t^2-x^2}-1\right)\right)^j}{\sqrt{t^2-x^2} j!}=-\frac{e^{\xi  \left(-\sqrt{x^2-t^2}\right)}}{4 \pi  \sqrt{x^2-t^2}}.
\end{equation}
It aligns perfectly with the bulk two-point function \eqref{eq: bulk 2pt} upon identification
\begin{equation}
    m=\xi.
\end{equation}
A similar discussion can be applied to the $y$-dependent part, which will not be repeated here. Combining the results, we have
\begin{equation}\label{eq: two-point function}
   \langle \phi(0,0,0)|\phi(t,x,y)\rangle=\frac{e^{i\xi}}{4\pi}\ ^\vee\!\langle 0|\phi(t,x,y)\rangle=-\frac{e^{-\xi D_{\text{flat}}}}{4\pi D_{\text{flat}}}.
\end{equation}
This is one of the key findings in this paper. It verifies that our bulk reconstruction of the local scalar excitation, as described in equations \eqref{eq: t solution} and \eqref{eq: general solution}, is valid. Additionally, it confirms that the appropriate bra state is indeed the $^\vee\langle0|$ state in the dual basis, as \eqref{eq: relation between conjugate state}. Furthermore, it offers the option to choose the sign as $\xi=m$, which provides a clear one-to-one correspondence between the bulk massive excitation spectrum and the CCFT$_2$ induced module.
\paragraph{Information metric} We will investigate the bulk geometry using the scalar excitation $|\phi(t,x,y)\rangle$ and determine that the information metric is proportional to the flat metric, extending the findings for AdS$_3$ \cite{Miyaji:2015fia} and dS$_3$ \cite{Doi:2024nty}.

The bulk two-point function, reproducible by the inner product \eqref{eq: two-point function}, exhibits a short-distance divergence of $\frac{1}{\epsilon}$, where $\epsilon$ has the dimension of length. This suggests that the state $|\phi(t,x,y)\rangle$ is singular. To regularize this, we can introduce
\begin{equation}
    |\phi(t,x,y\rangle)\to e^{\epsilon M_0}|\phi(t,x,y)\rangle.
\end{equation}
Equivalently, it gives a small imaginary part $i\epsilon$ to the time $t$. Then the two point function is
\begin{equation}
     \langle \phi(0,0,0)|\phi(t,x,y)\rangle=-\frac{e^{-\xi \tilde D_{\text{flat}}}}{4\pi \tilde D_{\text{flat}}},
\end{equation}
where
\begin{equation}
    \tilde D=\sqrt{-t^2+x^2+y^2+\epsilon^2+2t\epsilon}.
\end{equation}
So the short distance divergence can be reproduced,
\begin{equation}
    \lim_{t\to0,x\to0,y\to0}\langle \phi(0,0,0)|\phi(t,x,y)\rangle=-\frac{1}{4\pi\epsilon}.
\end{equation}
With the regulator parameter $\epsilon$, the inner product above can be further normalized to $1$ and has the short distance behaviour,
\begin{equation}
     \lim_{t\to0,x\to0,y\to0}\langle \phi(0,0,0)|\phi(t,x,y)\rangle\sim\frac{\epsilon}{\sqrt{-t^2+x^2+y^2+\epsilon^2+2t\epsilon}}.
\end{equation}

The information metric is defined by
\begin{equation}
    G_{ij}dx^idx^j=1-|\langle\phi(x)|\phi(x+dx)\rangle|,
\end{equation}
with the normalization
\begin{equation}
    \langle\phi(x)|\phi(x)\rangle=1.
\end{equation}
In the short-distance limit $x\to x'$,
we can expand $\langle\phi(x)|\phi(x')\rangle$ to have
\begin{equation}
    ds^2=\frac{1}{2\epsilon^2}D^2(x,x+dx),
\end{equation}
where
\begin{equation}
    D^2(x,x+dx)=-dt^2+dx^2+dy^2,
\end{equation}
is the flat distance. This then reproduces the metric of flat$_3$ spacetime from the scalar excitation in terms of the induced states \eqref{eq: t solution}.

\subsection{Solution in the highest weight representation}
\label{section: highest weight solution}
In the massless case, the highest weight representation solution is permissible, as the Casimir operators possess the following eigenvalues.
\begin{equation}\label{eq: highest weight casimir}
C_1|\Delta,\xi\rangle=\xi^2|\Delta,\xi\rangle,\ \ C_2|\Delta,\xi\rangle=2\xi(\Delta-1)|\Delta,\xi\rangle.
\end{equation}
In the bulk, the eigenvalue of $C_1$ vanishes, leading to
\begin{equation}
    \xi=m=0,
\end{equation}
and it solves the condition that the eigenvalue of $C_2$ vanishes.

Considering the excitation at origin $|\phi(0,0,0)\rangle$, it satisfies
\begin{equation}\label{3dconstra}
L_{0,\pm1}|\phi(0,0,0)\rangle=0.
\end{equation}
Assuming that we can construct this solution with a single module in the highest weight representation, the ansatz is
\begin{equation}
|\phi(0,0,0)\rangle=\sum_{i,j}c_{ij}L_{-1}^iM_{-1}^j|\Delta,\xi\rangle,
\end{equation}
where $|\Delta,\xi\rangle$ is a primary state. From the condition of action under $L_0$,
\begin{equation}
    L_0L_{-1}^iM_{-1}^j|\Delta,\xi\rangle=(\Delta+i+j)L_{-1}^iM_{-1}^j|\Delta,\xi\rangle,
\end{equation}
we can find the condition
\begin{equation}
    \Delta=-i-j.
\end{equation}
In our ansatz, $i,j$ are the numbers of the creation operator which are nonnegative integers, so that we should consider the negative integer $\Delta$. Then, we can express our ansatz as
\begin{equation}
|\phi(0,0,0)\rangle=\sum_{i=0}^{-\Delta}c_{i}L_{-1}^iM_{-1}^{-\Delta-i}|\Delta,\xi\rangle.
\end{equation}
From the following commutation relations,
\begin{equation}
    [L_1,L_{-1}^n]=2nL_{-1}^{n-1}L_0+n(n-1)L_{-1}^{n-1},
\end{equation}
\begin{equation}
    [L_1,M_{-1}^n]=2nM_{-1}^{n-1}M_0,
\end{equation}
\begin{equation}
    [L_0,M_{-1}^n]=nM_{-1}^n,
\end{equation}
we calculate the action under $L_0$ as follows
\begin{equation}
L_1L_{-1}^iM_{-1}^j|\Delta,\xi\rangle=(i(i-1+2\Delta+2j)L_{-1}^{i-1}M_{-1}^j+2j\xi L_{-1}^iM_{-1}^{j-1})|\Delta,\xi\rangle,
\end{equation}
which will in turn give the recursive relation
\begin{equation}
    c_iK_2(i,-\Delta-i)+c_{i+1}K_1(i+1,-\Delta-i-1)=0,\ \ c_0=1,
\end{equation}
where the functions $K_{1,2}$ are
\begin{equation}
    K_1(i,j)=i(i-1+2\Delta+2j),\ \ K_2(i,j)=2j\xi.
\end{equation}
The solution to this recursive relation reads
\begin{equation}
c_i=\frac{2^{i-1}\Delta(-\xi)^i(\Delta+1)_{i-1}}{(2)_{i-1}(3)_{i-1}}.
\end{equation}
For example, in the case where $\Delta=-1$, we can find the following solution,
\begin{equation}
|\phi(0,0,0)\rangle=M_{-1}|\Delta,\xi\rangle+\xi L_{-1}|\Delta,\xi\rangle.
\end{equation}
Further calculation shows that its norm is $2\xi^2$. As $\xi\rightarrow0$, it becomes a null state, although we should normalize it to $1$. The norm for arbitrary $\Delta$ is
\begin{equation}
    n_\Delta=\frac{(2\xi)^{-2\Delta}}{-\Delta+1}.
\end{equation}
Thus, the normalized coefficient is
\begin{equation}\label{eq: hwr solution}
    c_i=\frac{\sqrt{-\Delta+1}\Delta(2\xi)^{\Delta+i}(-1)^i(\Delta+1)_{i-1}}{2(2)_{i-1}(3)_{i-1}}.
\end{equation}
Then, what is left is the action under $L_{-1}$, which gives directly
\begin{equation}\label{eq:hwrlm}
    L_{-1}|\phi(0,0,0)\rangle=\sum c_iL_{-1}^{i+1}M_{-1}^{-\Delta-i}|\Delta,\xi\rangle.
\end{equation}
To see how this condition is satisfied, we calculate its inner product with $L_{-1}^{-\Delta+1}|\Delta,\xi\rangle$ as
\begin{equation}
    \langle \Delta,\xi|L_1^{-\Delta+1}L_{-1}|\phi(0,0,0)\rangle=\frac
    {\sqrt{-\Delta+1}}{\Gamma(\Delta)}.
\end{equation}
The gamma function's singularity in the denominator causes it to vanish for negative integer $\Delta$. Additionally, with $\xi=0$, the states $M_{-1}^n|\Delta,\xi\rangle$ are null. Consequently, the inner products of $L_{-1}|\phi(0,0,0\rangle)$ with all states in the theory are zero, confirming that $c_i$ is a solution to equation \eqref{eq:hwrlm} in the $\xi\rightarrow 0$ limit.

It is important to note that the solution includes terms such as $\frac{1}{\xi^n}M_{-1}^n|\Delta,\xi\rangle$, which remains non-zero as $\xi$ approaches zero. Therefore, $\xi$ serves as a necessary regulator, which we ultimately set to zero.

In \cite{Chen:2023naw}, the authors address the following conditions in the Poincar\'{e} patch, using a similar setup to identify the holographic dual of massless scalar excitations in the bulk, expressed through the highest weight representation
\begin{equation}
L_{-1}+rM_0|\phi\rangle=L_0+2rM_1|\phi\rangle= L_1|\phi\rangle=0.
\end{equation}
Their solution reads
    \begin{equation}
        |\phi(0,r,0)\rangle=\sum r^{-(\Delta+i+j)}\lambda_{i,j}L_{-1}^iM_{-1}^j|\Delta,\xi\rangle,
    \end{equation}
    where
    \begin{equation}
        \lambda_{i,j}=\frac{2^{-j}\xi^{-2j}(-1)^i\xi^{-i}\Gamma(i+j+\Delta)\Gamma(i+2j+2\Delta-1)}{\Gamma(i+1)\Gamma(j+1)\Gamma(\Delta)\Gamma(i+j+2\Delta-1)}.
    \end{equation}
To have a well-defined solution at origin, we must have
    \begin{equation}
        \Delta+i+j=0,
    \end{equation}
 so that there is a truncation with a negative integer $\Delta$. In this scenario, the coefficient  $\lambda_{i,j}$
    \begin{equation}
        \lambda_{i,-\Delta-j}=\sqrt{1-\Delta}\xi^\Delta\Gamma(1-\Delta)c_i,
    \end{equation}
is consistent with $c_i$ up to an overall constant. Therefore, in the highest weight representation framework, our solution \eqref{eq: hwr solution} aligns with the findings in \cite{Chen:2023naw}.

We end this subsection with the following comments on the highest weight solution. In the massless case, for each negative integer $\Delta$, there is a corresponding solution. When comparing the state space of CCFT with that of flat gravity, it appears that a single massless scalar excitation in the bulk corresponds to multiple states in the field theory, leading to difficulty in matching the spectrum.  On the other hand, from the eigenvalues \eqref{eq: highest weight casimir}, the potential dictionary for the massive scalar is $\xi^2=m^2,\Delta=1$. However, we have not found the solution in this case, by solving the equations \eqref{eq: origin equation} and \eqref{eq: t equation} directly. Besides, in the later discussion in the next sections, we will see that it it natural to derive the induced representation from the flat limit. This suggests that the highest weight representation is not an appropriate candidate for bulk scalar reconstruction.

\section{Flat limit from AdS$_3$/CFT$_2$}\label{sec:flat_limit_AdS}
        Discussion in the previous section suggests that the physically sensible representation of BMS$_3$ algebra \eqref{eq: bms_3 algebra} is not the highest weight representation \eqref{eq: bms highest weight rep} but the induced representation \eqref{eq: bms induced rep}.
        This is supported by the fact that the induced representation in Flat$_3$/CCFT$_2$ is indeed a certain limit of---or, say, ``induced" from (as its name suggest)---the highest weight representation in AdS$_3$/CFT$_2$ as we shortly check below (not to be confused with the highest weight representation of BMS$_3$).
        \par The main concern in this section is to show that not only the algebra generators but also the bulk local scalar state and the Green's function are in truth ``induced" from that in AdS$_3$/CFT$_2$.
        To see this, we first review how the AdS$_3$/CFT$_2$ Virasoro algebra gives rise to the BMS$_3$ algebra in the flat limit in Subsection \ref{subsec: virasoro and highest weight}.
        The flat limit, or sometimes called the \textit{ultra-relativistic} (\textit{UV}) \textit{limit}, of AdS$_3$ amounts to taking $l\to\infty$ because the AdS$_3$ metric in the global coordinate
        \begin{align}
            g_{\text{AdS}_3}=-\Big(1+\frac{r^2}{l^2}\Big)dt^2+\frac{1}{1+\frac{r^2}{l^2}}dr^2+r^2d\phi^2,   \label{eq: global coordinate of AdS}
        \end{align}
        asymptotes to the flat$_3$ metric in the polar coordinate \eqref{eq: flat polar coordinate}.
        We can alternatively see this by coordinate transformation $u:=t-l\operatorname{arctan}\frac{r}{l}$ and move to the metric \eqref{eq: AdS metric in u,r,phi}, whose $l\to\infty$ limit coincides with the flat$_3$ metric \eqref{eq: 3d flat metric in the retarded coordinate} in retarded coordinate.
        We then proceed to prove that the bulk local state in AdS$_3$/CFT$_2$ \cite{Miyaji:2015fia} correctly connects to that in Flat$_3$/CCFT$_2$ in the flat limit $l\to\infty$ in Subsection \ref{subsec: states_correspondence_AdS}.
        The analysis starts with verifying that the constraint equations on AdS$_3$ bulk local states agree with that on the flat bulk local states as $l\to\infty$, and then we prove that the states on both sides are linked.

        \subsection{Flat limit of AdS$_3$ Virasoro algebra and the highest weight representation} \label{subsec: virasoro and highest weight}
            To discuss its flat limit, let us first summarize the basic properties of the Virasoro algebra $\operatorname{Vir}\times\overline{\operatorname{Vir}}$ with $c_{\text{AdS}_3}=\overline{c}_{\text{AdS}_3}=\frac{3l}{2G}$ the Brown-Henneaux value in AdS$_3$/CFT$_2$ with the AdS radius $l$,
            \begin{align}
                \begin{aligned}
                    [\mathcal{L}_m,\mathcal{L}_n]&=(m-n)\mathcal{L}_{m+n}+\frac{c_{\text{AdS}_3}}{12}(m^3-m)\delta_{m+n,0}, \\
                    [\overline{\mathcal{L}}_m,\overline{\mathcal{L}}_n]&=(m-n)\overline{\mathcal{L}}_{m+n}+\frac{\overline{c}_{\text{AdS}_3}}{12}(m^3-m)\delta_{m+n,0}, \\
                    [\mathcal{L}_m,\overline{\mathcal{L}}_n]&=0,
                \end{aligned}
                \qquad(m,n\in\mathbb{Z}).
            \label{eq: Virasoro commutation relations}
            \end{align}
            We adopt the calligraphy $\mathcal{L}_n$ to denote the Virasoro generators for distinction from BMS$_3$ generators $L_n$.
            In AdS$_3$/CFT$_2$, the physical states are described using the highest weight representation (HWR) built upon an primary state $|h,\overline{h}\rangle:=\ket*{h}\otimes|\overline{h}\rangle$ satisfying
            \begin{align}
                \mathcal{L}_0|h,\overline{h}\rangle=h|h,\overline{h}\rangle,\:\:\overline{\mathcal{L}}_0|h,\overline{h}\rangle=\overline{h}\:|h,\overline{h}\rangle,\:\mathcal{L}_n|h,\overline{h}\rangle=\overline{\mathcal{L}}_n|h,\overline{h}\rangle=0\:(n>0). \label{eq: highest weight condition}
            \end{align}
            The total Hilbert space $\mathcal{H}$ of the boundary CFT$_2$ is spanned by the set of descendants
            \begin{align}
                \mathcal{L}_{-n_1}\cdots\mathcal{L}_{-n_k}\overline{\mathcal{L}}_{-\overline{n}_1}\cdots\overline{\mathcal{L}}_{-\overline{n}_l}|h,\overline{h}\rangle\quad(n_1\geq \cdots\geq n_k\geq 0,\,\overline{n}_1\geq \cdots\geq \overline{n}_l\geq 0),
            \end{align}
            and in order for the representation to be unitary (\,all state vectors $\ket*{\phi}\in\mathcal{H}$ have non-negative norm), the Virasoro generators must obey the conjugate rules
            \begin{align}
                \mathcal{L}_n^{\dagger}=\mathcal{L}_{-n},\:\overline{\mathcal{L}}_n^{\dagger}=\overline{\mathcal{L}}_{-n}. \label{eq: Virasoro conjugation rule}
            \end{align}
            Then, let us apply a suitable procedure to reproduce the BMS$_3$ algebra from the AdS$_3$ Virasoro algebra \eqref{eq: Virasoro commutation relations} in its flat limit $l\to\infty$.
            To this end, we define new generators\footnote{
                This is different from the conventional one $L_{n}^{\text{conv.}}(l):=\mathcal{L}_n-\overline{ \mathcal{L}}_{-n},\;M_{n}^{\text{conv.}}(l):=\frac{1}{l}(\mathcal{L}_n+\overline{ \mathcal{L}}_{-n})$ \cite{Barnich:2012aw, Campoleoni:2016vsh, Bagchi:2017cpu}.
                Indeed, if we assume that the correct linear combinations that reproduce eqs.\,\eqref{eq: BMS-like commutators before the flat limit} \eqref{eq: BMS-like central charge}, and \eqref{eq: flat conjugation before flat limit} are of the form
                \begin{align}
                    \begin{aligned}
                        L_n(l) & =a_n\mathcal{L}_n+b_n\mathcal{L}_{-n}+c_n\overline{\mathcal{L}}_n+d_n\overline{\mathcal{L}}_{-n} & (a_n,b_n,c_n,d_n\in\mathbb{C}), \\
                        M_n(l) & =\alpha_n\mathcal{L}_n+\beta_n\mathcal{L}_{-n}+\gamma_n\overline{\mathcal{L}}_n+\delta_n\overline{\mathcal{L}}_{-n} & (\alpha_n,\gamma_n,\delta_n,\delta_n\in\mathbb{C}),
                    \end{aligned}
                \end{align}
                then there remains an ambiguity up to some constants.
                More precisely, all of the following linear combination are the solution to the required conditions:
                \begin{gather}
                    L_{n}(l):=e^{inx_1}\mathcal{L}_n-e^{iny_1}\overline{\mathcal{L}}_{-n},\;\;
                    M_{n}(l):=\frac{1}{l}(e^{inx}\mathcal{L}_n+e^{iny}\overline{ \mathcal{L}}_{-n})\qquad\big(x_1,y_1\in[0,2\pi)\big) \\
                    \text{or} \notag \\
                    L_{n}(l):=e^{inx_2}\mathcal{L}_{-n}-e^{iny_2}\overline{\mathcal{L}}_n,\;\;
                    M_{n}(l):=\frac{1}{l}(e^{inx_2}\mathcal{L}_{-n}+e^{iny_2}\overline{ \mathcal{L}}_n)\qquad\big(x_2,y_2\in[0,2\pi)\big).
                \end{gather}
                However, we can fix the value of $(x_i,y_i)$ by demanding that the dual bulk isometries also matches to \eqref{eq: flat3killing} in the flat limit (See Appendix \ref{append: AdS generators} for explicit form of isometries in AdS$_3$).
                The answer is $(x_1,y_1)=(0,\pi)$, rather than $(x_1,y_1)=(0,0)$, in our convention.
            }
            \begin{align}
                 L_{n}(l):=\mathcal{L}_n-(-1)^n\overline{ \mathcal{L}}_{-n},\;\;
                 M_{n}(l):=\frac{1}{l}(\mathcal{L}_n+(-1)^n\overline{ \mathcal{L}}_{-n}),   \label{eq: generator relation between flat and AdS}
            \end{align}
            which succeeds in replicating the BMS-like commutators
            \begin{align}
                \begin{aligned}
                    [L_m(l),L_n(l)] & =(m-n)L_{m+n}(l)+\frac{c_L}{12}(m^3-m)\delta_{m+n,0}, \\
                    [L_m(l),M_n(l)] & =(m-n)M_{m+n}(l)+\frac{c_M}{12}(m^3-m)\delta_{m+n,0}, \\
                    [M_m(l),M_n(l)] & =\frac{1}{l^2}\Big((m-n)L_{m+n}(l)+\frac{c_L}{12}(m^3-m)\delta_{m+n,0}\Big).
                \end{aligned} \label{eq: BMS-like commutators before the flat limit}
            \end{align}
            The new ``central charges" $c_L$, $c_M$ are related to the Virasoro central charges $c_{\text{AdS}_3}$, $\overline{c}_{\text{AdS}_3}$ by
            \begin{align}
                c_L:=c_{\text{AdS}_3}-\overline{c}_{\text{AdS}_3}=0,\quad c_M:=\frac{c_{\text{AdS}_3}+\overline{c}_{\text{AdS}_3}}{l}=\frac{3}{G},  \label{eq: BMS-like central charge}
            \end{align}
            which are totally identical to the BMS central charges \eqref{eq: BMS central charge}.
            It also follows from eq.\,\eqref{eq: Virasoro conjugation rule} that eq.\,\eqref{eq: generator relation between flat and AdS} properly satisfies the conjugation rule
            \begin{align}
                L_n(l)^{\dagger}=L_{-n}(l),\;\; M_n(l)^{\dagger}=M_{-n}(l). \label{eq: flat conjugation before flat limit}
            \end{align}
            By taking $l\to\infty$, the commutators \eqref{eq: BMS-like commutators before the flat limit} indeed match with the BMS$_3$ algebra \eqref{eq: bms_3 algebra} with the scalar value \eqref{eq: conjugation relation}.
            \par We now go on to confirm that the highest weight representation turns into the induced representation in the flat limit.
            One readily observes this when we rewrite the highest weight conditions \eqref{eq: highest weight condition} in terms of the new generators \eqref{eq: generator relation between flat and AdS}
            \begin{gather}
                L_0(l)|h,\overline{h}\rangle=(h-\overline{h})|h,\overline{h}\rangle,\:\:M_0(l)|h,\overline{h}\rangle=\frac{h+\overline{h}}{l}|h,\overline{h}\rangle, \\
                \Big(M_n(l)+\frac{1}{l}L_n(l)\Big)|h,\overline{h}\rangle=0,\:\:\Big(M_{-n}(l)-\frac{1}{l}L_{-n}(l)\Big)|h,\overline{h}\rangle=0\quad(n>0). \label{eq: highest weight condition in AdS represented in flat generators}
            \end{gather}
            Using the variables $\displaystyle\xi:=\lim_{l\to\infty}\frac{1}{l}(h+\overline{h}),\;\;\Delta:=\lim_{l\to\infty}(h-\overline{h})$,
            these conditions result in the induced conditions
            \begin{align}
                M_0|\Delta,\xi\rangle=\xi|\Delta,\xi\rangle,\:M_n|\Delta,\xi\rangle=0\,(n\neq 0),\:L_0|\Delta,\xi\rangle=\Delta|\Delta,\xi\rangle,  \label{eq: BMS induced from AdS highest}
            \end{align}
            where $\displaystyle L_n:=\lim_{l\to\infty}L_n(l),\;M_n:=\lim_{l\to\infty}M_n(l),\;|\Delta,\xi\rangle:=\lim_{l\to\infty}|h,\overline{h}\rangle$.
            Since we restrict our attention to scalar field $h=\overline{h}=\frac{1}{2}+\frac{1}{2}\sqrt{m^2l^2+1}$ in this paper, the limit value $\xi=m, \Delta=0$ is convergent.
            In this case, eq.\,\eqref{eq: BMS induced from AdS highest} is in complete agreement with the scalar BMS$_3$ induced condition \eqref{eq: bms induced rep} with eq.\,\eqref{eq: xi and weight for scalar in bms}.
\subsection{Casimir}
    Let us briefly check whether we can reproduce Casimir operators from the one in the AdS space. In the AdS, the scalar Casimir operator $\Box=1/\sqrt{-g}\partial_\mu(\sqrt{-g}g^{\mu\nu}\partial_\nu)$ is rewritten by  the left and right Virasoro generators as
\begin{equation}
\Box_L=\frac{\mathcal{L}_{-1}\mathcal{L}_{1}+\mathcal{L}_{1}\mathcal{L}_{-1}}{2}-\mathcal{L}_{0}^2,\quad \Box_R=\frac{\bar{\mathcal{L}}_{-1}\bar{\mathcal{L}}_{1}+\bar{\mathcal{L}}_{1}\bar{\mathcal{L}}_{-1}}{2}-\bar{\mathcal{L}}_{0}^2.
\end{equation}
We can rewrite the Virasoro generators in the AdS to those in the flat case. Then, we can extract the $C_1$ as
\begin{align}
    \Box_L+\Box_R&\sim\frac{l^2M_{-1}M_1}{4}-\frac{l^2M_0^2}{2}\nonumber\\
    &=-\frac{l^2}{2}C_1,
\end{align}
where we pick up $l^2$ terms in the first line since it is the leading contribution.
Also, the $C_2$ can be derived directly as
\begin{align}
    \Box_L-\Box_R&\sim l\frac{M_{-1}L_1+L_1M_{-1}+M_1L_{-1}+L_{-1}M_1}{4}-lM_0L_0\nonumber\\
    &=-\frac{l}{2}C_2,
\end{align}
where again we pick up terms proportional to $l$ in the leading order.

     \subsection{Bulk local state correspondence} \label{subsec: states_correspondence_AdS}
            From now on, we only discuss the global part $\{\mathcal{L}_n,\overline{\mathcal{L}}_n\}_{n=0,\pm 1}$ of the Virasoro algebra \eqref{eq: Virasoro commutation relations} as in the Flat$_3$/BMS$_2$ case so that we neglect the effect of the boundary gravitons (the descendants in the vacuum Verma module) which correspond to the metric fluctuation in the bulk.
            This means we take no account of the quantum fluctuation of spacetime at this time.
            In AdS$_3$/CFT$_2$, the CFT state $\ket*{\Psi_\text{AdS}(t=0,r=0)}$ dual to the bulk local scalar state at the origin $(t,\theta)=(0,0)$ is supposed to satisfy the constraint
            \begin{align}
                (\mathcal{L}_0-\overline{\mathcal{L}}_0)\ket*{\Psi_\text{AdS}(t=0,r=0)}=0,\;\;(\mathcal{L}_{\pm 1}+\overline{\mathcal{L}}_{\mp 1})\ket*{\Psi_\text{AdS}(t=0,r=0)}=0,  \label{eq: the constraint for the excitation at origin in AdS}
            \end{align}
            because $\mathcal{L}_0-\overline{\mathcal{L}}_0$, $\mathcal{L}_{\pm 1}+\overline{\mathcal{L}}_{\mp 1}$ are the generators of the stabilizer subgroup of the origin in the bulk (see Appendix \ref{append: AdS generators} for explicit form of AdS$_3$ isometry generators).
            The solution is given by \cite{Miyaji:2015fia}
            \begin{align}
                \ket*{\Psi_{\text{AdS}}(t=0,r=0)}=e^{\frac{i \pi}{2}(\mathcal{L}_0+\overline{\mathcal{L}}_0-2h)}|I_h\rangle,
            \end{align}
            starting from the \textit{Ishibashi state}\footnote{
                The Ishibashi state $|I_h\rangle$ is defined exclusively in diagonal CFT and satisfies
                \begin{align}
                    (\mathcal{L}_n-\overline{\mathcal{L}}_{-n})|I_h\rangle=0\quad(\forall n\in\mathbb{Z}).
                \end{align}
                It is used to construct boundary states in BCFT \cite{Cardy:2004hm}
                \begin{align}
                    |B\rangle\!\rangle=\sum_{h}c_h|I_h\rangle\quad(c_h\in\mathbb{C}).
                \end{align}
            }
            \begin{align}
                |I_h\rangle:=\sum_{k=0}^{\infty}|k,k\rangle,    \label{eq: AdS  Ishibashi state}
            \end{align}
            where each summand is the normalized Virasoro descendant
            \begin{align}
                \ket*{k,k}:=\Bigg(\prod_{j=1}^k\frac{1}{j\big(2h-1+j\big)}\Bigg)\mathcal{L}_{-1}^k\overline{\mathcal{L}}_{-1}^k|h,h\rangle.
            \end{align}
            The excitation at the general point $(t,r,\phi)$ is given by
            \begin{align}
                \ket*{\Psi_{\text{AdS}}(t,r,\phi)}=e^{-i(\mathcal{L}_0+\overline{\mathcal{L}}_0)\frac{t}{l}}e^{\frac{1}{2}(\mathcal{L}_1-\overline{\mathcal{L}}_1-\mathcal{L}_{-1}+\overline{\mathcal{L}}_{-1})\operatorname{arcsinh}\frac{r}{l}}e^{i(\mathcal{L}_0-\overline{\mathcal{L}}_0)\phi}\ket*{\Psi_{\text{AdS}}(t=0,r=0)}.
            \end{align}
            \par Next, we investigate the coordinates correspondence between AdS$_3$ and flat$_3$, as well as the relation between AdS$_3$ constraints and the flat$_3$ ones.

            \subsubsection*{(1) AdS$_3$ origin $(t,r)=(0,0)$ $\to$ Flat$_3$ origin $(t,r)=(0,0)$}
                First, the AdS origin $(t,r)=(0,0)$ maps directly to the flat$_3$ origin $(t,r)=(0,0)$ in the flat limit $l\to\infty$ as is evident from the global coordinate metric \eqref{eq: global coordinate of AdS}.
                In fact, the generic AdS$_3$ point $(t,r,\phi)$ maps to the point $(t,r,\phi)$ in flat$_3$.
                If we assume the existence of a putative limit state $\displaystyle\ket*{\phi(t,r,\phi)}:=\lim_{ l\to\infty}\ket*{\Psi_{\text{AdS}}(t,r,\phi)}$ (which will be given later), the constraint \eqref{eq: the constraint for the excitation at origin in AdS} leads in the flat limit to
                \begin{align}
                    L_0\ket*{\phi(t=0,r=0)}=0,\;\;L_{\pm 1}\ket*{\phi(t=0,r=0)}=0,
                \end{align}
                as one can see from eq.\,\eqref{eq: generator relation between flat and AdS}.
                This is the rotational and boost invariance at origin \eqref{eq: origin equation} in the flat$_3$.

            \subsubsection*{(2) AdS$_3$ $(t,r,\phi)=(t,0,0)$ $\to$ Flat$_3$ $(t,r,\phi)=(t,0,0)$}
            For clarity, we also discuss the constraints that $\ket*{\Psi_\text{AdS}(t,0,0)}$ must follow at the point shifted from the origin by a constant time $t$.
            The time translation operator is $e^{i(\mathcal{L}_0+\overline{\mathcal{L}}_0)\frac{t}{l}}$, so the constraints at the origin \eqref{eq: the constraint for the excitation at origin in AdS} are transformed into
            \begin{align}
                (\mathcal{L}_0-\overline{\mathcal{L}}_0)\ket*{\Psi_\text{AdS}(t,0,0)}=0,\;\;(e^{\mp i\frac{t}{l}} \mathcal{L}_{\pm 1}+e^{\pm i\frac{t}{l}}\overline{\mathcal{L}}_{\mp 1})\ket*{\Psi_\text{AdS}(t,0,0)}=0. \label{eq: AdS constraint at (t,0,0)}
            \end{align}
            Here, we use the identity $e^{ia\mathcal{L}_0}\mathcal{L}_{\pm 1} e^{-ia\mathcal{L}_0}=e^{\mp ia}\mathcal{L}_{\pm 1}\;\;(a\in\mathbb{R})$.
            Thus, the flat limit of these conditions are
            \begin{align}
                 L_0\ket*{\phi(t,0,0)}=0,\;\;(L_{\pm 1}\mp it M_{\pm 1})\ket*{\phi(t,0,0)}=0, \label{eq:UR_limit_of_constraint_from_AdS}
            \end{align}
            where $\displaystyle\ket*{\phi(t,r,\phi)}$ is the state obtained by the  flat limit.
            The second equations result from eq.\,\eqref{eq: generator relation between flat and AdS}
            \begin{align}
                \lim_{l\to\infty}(e^{\mp i\frac{t}{l}} \mathcal{L}_{\pm 1}+e^{\pm i\frac{t}{l}}\overline{\mathcal{L}}_{\mp 1}) & =
                \lim_{l\to\infty}\left[\left(1\mp i \frac{t}{l}\right)\mathcal{L}_{\pm 1}+\left(1\pm i \frac{t}{l}\right)\overline{\mathcal{L}}_{\mp 1}+O\left(\frac{1}{l^2}\right)\right] \nonumber \\[1ex]
                & =L_{\pm 1}\mp it M_{\pm 1}.
            \end{align}
            Comparing  eq.\,\eqref{eq:UR_limit_of_constraint_from_AdS} to  eq.\,\eqref{eq: t equation}, we confirm that the flat limit appropriately connects the AdS$_3$ constraints to flat$_3$ ones, for which we expect that the the bulk local states in AdS$_3$ $\ket*{\Psi_\text{AdS}(t,0,0)}$ and that of Flat$_3$ $\ket*{\Psi_\text{Flat}(t,0,0)}$ be connected via the flat limit, i.e.\,$\displaystyle \ket*{\phi(t,0,0)}=\lim_{l\to\infty}\ket*{\Psi_\text{AdS}(t,0,0)}=\ket*{\Psi_\text{Flat}(t,0,0)}$.
            We will shortly observe that this is indeed the case modulo a constant factor depending on $l$.
            \ \\
            \par The solution to eq.\,\eqref{eq: AdS constraint at (t,0,0)} is
            \begin{align}
                \begin{aligned}
                    \ket*{\Psi_\text{AdS}(t,0,0)} & =e^{i\frac{t}{l}(\mathcal{L}_0+\overline{\mathcal{L}}_0-2h)}\ket*{\Psi_\text{AdS}(0,0,0)} \\
                    & =e^{i(\frac{\pi}{2}+\frac{t}{l})(\mathcal{L}_0+\overline{\mathcal{L}}_0-2h)}|I_h\rangle=\sum_{k=0}^\infty \frac{e^{ik(\pi+\frac{2t}{l})}}{\Gamma (k+1) (2h)_k}\mathcal{L}_{-1}^k\overline{\mathcal{L}}_{-1}^k|h,h\rangle,
                \end{aligned}
                \label{eq: AdS bulk local state at (t,0,0)}
            \end{align}
            where $|I_h\rangle$ is the Ishibashi state \eqref{eq: AdS  Ishibashi state}.
            After some  algebra, we  rewrite the AdS$_3$/CFT$_2$ descendants in terms of ``flat descendants" built by acting $L_{-1}(l)$ and $L_1(l)$,
            \begin{align}
                \mathcal{L}_{-1}^k\overline{\mathcal{L}}_{-1}^k|h,h\rangle=\sum_{a=0}^k f_{k,a}\big(L_{-1}(l)\big)^a\big(L_{1}(l)\big)^a|h,h\rangle, \label{eq: AdS descendants and fla descendants}
            \end{align}
            where the coefficient is given by
            \begin{align}   \label{eq: AdS flat descendant coefficient}
               f_{k,a}= \frac{(-1)^{k-a} \Gamma (k+1)^2 \Gamma (k+2h)}{\Gamma (a+1)^2 \Gamma (-a+k+1) \Gamma (a+2h)}.
            \end{align}
            See Appendix \ref{append: proof of descendant relation} for the complete proof.
            Then, the AdS bulk local state at $(t,0,0)$ \eqref{eq: AdS bulk local state at (t,0,0)} can be expanded in terms of the flat descendants
            \begin{align}
                 \ket*{\Psi_\text{AdS}(t,0,0)}=\sum_{a=0}^\infty\underbrace{\sum_{k=0}^\infty\frac{e^{ik(\pi+\frac{2t}{l})}}{\Gamma (k+1) (2h)_k}f_{k,a}}_{=:\,f_a}\big(L_{-1}(l)\big)^{a}\big(L_{1}(l)\big)^{a}|h,h\rangle.
            \end{align}
            To see if this state asymptotes to $|\Psi_{\text{Flat}}(t,0,0)\rangle$ as $l\to\infty$ one needs to keep track of the coeficient
            \begin{equation}
                f_a=\sum_{k=0}^\infty \frac{e^{ik(\pi+2\frac{t}{l})}}{\Gamma (k+1) (2h)_k}f_{k,a}=\frac{(-1)^{-a} \, _2\tilde{F}_1\big(1,1;1-a;e^{\frac{2it}{l}}\big)}{\Gamma (a+1)^2 (2h)_a},
            \end{equation}
            where
            \begin{equation}
                _2\tilde{F}_1\big(p,q;r;z\big):=\frac{_2F_1\big(p,q;r;z\big)}{\Gamma(r)}=\sum_{n=0}^{\infty}\frac{(p)_n(q)_n}{\Gamma(r+n)}\frac{z^n}{n!}, \label{eq: regularized hypergeometric function}
            \end{equation}
            is the \textit{regularized hypergeometric function}.
            One should note that $\Gamma(z+n)=(z+n-1)(z+n-2)\cdots z\Gamma(z)=(z)_n\Gamma(z)$ for $n\in\mathbb{N}$.
            As $l\to\infty$, it becomes\footnote{
                In computing the first equality, it is worth noting that $\Gamma(-a+k+1)$ is divergent for $k<a$ and
                \begin{align}
                    \frac{1}{(1-x)^{a+1}}=\sum_{k=0}^{\infty}\left(\begin{matrix}
                        k+a \\
                        a
                    \end{matrix}\right)
                    x^k\quad(|x|<1).
                \end{align}
            }
            \begin{equation}
                \lim_{l\to\infty}\frac{f_a}{l}=-\frac{2^{-a-1}m^{-a}(-\frac{i}{t})^{a+1}}{\Gamma (a+1)},\;\;\lim_{l\to\infty}\big(L_{-1}(l)\big)^{a}\big(L_{1}(l)\big)^{a}|h,h\rangle=L_{-1}^{a}L_{1}^{a}|0,\xi\rangle,
            \end{equation}
            thus we show
            \begin{equation}
            -\lim_{l\to\infty}\frac{2}{l}\ket*{\Psi_\text{AdS}(t,0,0)}=\sum_{a=0}^\infty\frac{2^{-a}m^{-a}(-\frac{i}{t})^{a+1}}{\Gamma (a+1)}L_{-1}^{a}L_{1}^{a}|0,\xi\rangle=\ket*{\Psi_\text{Flat}(t,0,0)}.
            \end{equation}
            At $t=0$, it is divergent when taking the limit because $ \, _2\tilde{F}_1(1,1;1-a,x)$ has the singularity at $x=1$. \footnote{The regularized hypergeometric function \eqref{eq: regularized hypergeometric function} diverges at $z=1$ if $\Re(r - p - q) \leq 0$ since the hypergeometric function
            \begin{equation}
                _2F_1(p,q;r;z):=\sum_{n=0}^{\infty}\frac{(p)_n(q)_n}{(r)_n}\frac{z^n}{n!}.
            \end{equation}
            diverges in that regime.}
            This is consistent with the fact that the solution \eqref{eq: t solution} is divergent at $t=0$.

\subsection{Green's functions correspondence}\label{greens_function_AdS}
As is shown in \cite{Miyaji:2015fia}, the Green's functions in AdS$_3$ are obtained by the inner product between  $ \ket*{\Psi_{\text{AdS}}(t,r,\phi)}$ and its Hermitian conjugate $( \ket*{\Psi_{\text{AdS}}(t,r,\phi)})^\dagger$:
\begin{equation}
    G_{\text{AdS}}(x_1,x_2)=\ev*{\Psi_{\text{AdS}}(t_1,r_1,\phi_1)|\Psi_{\text{AdS}}(t_2,r_2,\phi_2)}
    =\frac{e^{-\sqrt{m^2l^2+1}D_{\text{AdS}}/l}}{2\sinh D_{\text{AdS}}/l},
\end{equation}
where $D_{\text{AdS}}(x_1,x_2)$ is the geodesic length
\begin{equation}
    \cosh \frac{D_{\text{AdS}}(x_1,x_2)}{l}=\cos(\tau_1-\tau_2)\cosh\rho_1\cosh\rho_2-\cos(\phi_1-\phi_2)\sinh\rho_1\sinh\rho_2,
\end{equation}
expressed in terms of the global coordinates with a metric
\begin{equation}
ds^2=l^2(-\cosh^2\rho d\tau^2+d\rho^2+\sinh ^2\rho d\phi^2 ).
\end{equation}
The relation to the original global coordinates with a metric
\begin{equation}
      ds^2=-\Big(1+\frac{r^2}{l^2}\Big)dt^2+\frac{1}{1+\frac{r^2}{l^2}}dr^2+r^2d\phi^2,
\end{equation}
is given by
\begin{equation}
    \tau:=\frac{t}{l}, \quad \rho:=\operatorname{arcsinh}\frac{r}{l},\quad \phi:=\phi.
\end{equation}
Under the UR limit, $l\rightarrow \infty$, the Green's function reduces to
\begin{equation}
    \lim_{l\to\infty}\frac{1}{l}G_{\text{AdS}}(x_1,x_2)=\frac{e^{\pm m D_{\text{flat}} }}{2 D_{\text{flat}}}=G_{\text{flat}}(x_1,x_2),
\end{equation}
since the flat limit of geodesic length gives
\begin{equation}
    \lim_{l\to\infty}D_{\text{AdS}}(x_1,x_2)^2=D_{\text{flat}}(x_1,x_2)^2=-t^2+r^2.
\end{equation}
Here, the factor $l^{-1}$ is due to the additional $l^2$ in the AdS Laplacian compared to the flat one.

The flat limit of the bra state for scalar excitations on AdS$_3$ is expected to coincide with the conjugate of the induced representation introduced in Section \ref{section: induced solution}, but as mentioned there, its inner product cannot be computed directly. However, the flat limit of the AdS$_3$ inner product (i.e.\ the Green’s function) correctly reproduces the Green’s function of the  flat$_3$, thereby validating this method. \footnote{We also comment about the flat limit of bra states in section \ref{section: conclusion and outlook}.}

    \section{Flat limit from dS$_3$/CFT$_2$}\label{sec:flat_limit_dS}
        Inspired by the discussion of AdS$_3$/CFT$_2$ in the previous section, it is natural to infer that dS$_3$/CFT$_2$ likewise relates to Flat$_3$/CCFT$_2$ in its flat limit.
        In this case, the flat limit of dS$_3$ corresponds to setting the de Sitter radius $l$  to infinity because the dS$_3$ metric in the static patch coordinates
        \begin{align}
            g_{\text{dS}}=-\Big(1-\frac{r^2}{l^2}\Big)dt^2+\frac{1}{1-\frac{r^2}{l^2}}dr^2+r^2d\phi^2,
        \end{align}
        approaches the flat$_3$ metric expressed in the polar coordinate \eqref{eq: flat polar coordinate}.
        We can alternatively see this by coordinate transformation $u:=t-l\operatorname{arctanh}\frac{r}{l}$ and move to the metric \eqref{eq: dS metric in u,r,phi}, whose $l\to\infty$ limit coincides with the flat$_3$ metric \eqref{eq: 3d flat metric in the retarded coordinate} in the retarded coordinates.
        The dS/CFT correspondence is far less understood than AdS/CFT: debate persists over whether the boundary CFT$_2$ resides on the asymptotic spacelike boundary at temporal infinity \cite{Strominger:2001pn,Maldacena:2002vr}, on the cosmological horizon \cite{Susskind:2021esx,Susskind:2021omt} (including a timelike boundary \cite{Kawamoto:2023nki,Anninos:2023epi,Anninos:2024wpy,Silverstein:2024xnr,Noumi:2025lbb}), or at an observer \cite{Chandrasekaran:2022cip,Narovlansky:2023lfz,Verlinde:2024zrh}. In the first possibility, namely the original dS/CFT proposal, it has been known that the central charge of CFT$_2$ dual to dS$_3$ is imaginary \cite{Maldacena:2002vr,Hikida:2021ese} in the semiclassical limit  and is described by the Liouville CFT \cite{Hikida:2022ltr,Verlinde:2024zrh,Collier:2024kmo}. This raises the question how the bulk time emerges from the putative non-unitary Euclidean CFT$_2$.
        Despite these intricacies, we proceed to analyze the flat limit of dS$_3$/CFT$_2$ in a simple setup based on the previous work \cite{Doi:2024nty}.
        \par The punchline of this section comprises two primary steps: we first show that the induced representation in Flat$_3$/CCFT$_2$ is the consequence of the flat limit of the highest weight representation in dS$_3$/CFT$_2$ as in the AdS$_3$/CFT$_2$ case, and then show that the bulk local scalar state and the Green's function are ``induced" from that in dS$_3$/CFT$_2$.
        Subsection \ref{subsec: ds virasoro algebra and hwr} is devoted to find how the dS$_3$/CFT$_2$ Virasoro algebra gives rise to the BMS$_3$ algebra as $l\to\infty$. In subsection \ref{subsec: states_correspondence_dS}, we then confirm that the constraint equations on the dS$_3$/CFT$_2$ bulk local states reproduce that in Flat$_3$/CCFT$_2$, and eventually see that the states themselves are identical.
        The analysis is almost parallel to AdS$_3$/CFT$_2$ except the subtleties in bra states and the CPT gauging of the states \cite{Doi:2024nty} specific to dS$_3$/CFT$_2$.

        \subsection{Flat limit of dS$_3$ Virasoro algebra and the highest weight representation}    \label{subsec: ds virasoro algebra and hwr}
            Consider the Virasoro algebra $\operatorname{Vir}\times\overline{\operatorname{Vir}}$ with the Brown-Henneaux relation $c_{\text{dS}_3}=\overline{c}_{\text{dS}_3}=\frac{3l}{2G}i$ in the  dS$_3$/CFT$_2$ ($l$\,:\,dS radius),
            \begin{align}
                \begin{aligned}
                    [\mathcal{L}_m,\mathcal{L}_n]&=(m-n)\mathcal{L}_{m+n}+\frac{c_{\text{dS}_3}}{12}(m^3-m)\delta_{m+n,0}, \\
                    [\overline{\mathcal{L}}_m,\overline{\mathcal{L}}_n]&=(m-n)\overline{\mathcal{L}}_{m+n}+\frac{\overline{c}_{\text{dS}_3}}{12}(m^3-m)\delta_{m+n,0}, \\
                    [\mathcal{L}_m,\overline{\mathcal{L}}_n]&=0.
                \end{aligned}
                \qquad(m,n\in\mathbb{Z}).
                \label{eq: Virasoro commutation relations in dS}
            \end{align}
            We again employ the calligraphy $\mathcal{L}_n$ to denote the Virasoro generators for distinction from BMS$_3$ generators $L_n$.
            The physical states in dS$_3$/CFT$_2$ are also represented by the highest weight representation built upon a primary state $|h,\overline{h}\rangle:=\ket*{h}\otimes|\overline{h}\rangle$ satisfying
            \begin{align}
                \mathcal{L}_0|h,\overline{h}\rangle=h|h,\overline{h}\rangle,\:\:\overline{\mathcal{L}}_0|h,\overline{h}\rangle=\overline{h}\:|h,\overline{h}\rangle,\:\mathcal{L}_n|h,\overline{h}\rangle=\overline{\mathcal{L}}_n|h,\overline{h}\rangle=0\:(n>0). \label{eq: highest weight condition in dS}
            \end{align}
            based on which the Verma module $\mathcal{V}_{h,\overline{h}}$ of the boundary CFT$_2$ is constructed by acting the collection of Virasoro generators:
            \begin{align}
                \mathcal{V}_{h,\overline{h}}=\operatorname{span}\left\{\mathcal{L}_{-n_1}\cdots\mathcal{L}_{-n_k}\overline{\mathcal{L}}_{-\overline{n}_1}\cdots\overline{\mathcal{L}}_{-\overline{n}_l}|h,\overline{h}\rangle\middle|n_i,\overline{n}_i\in\mathbb{N}\cup\{0\}\right\},
            \end{align}
            The conjugation rule is non-standard as \cite{Doi:2024nty}
            \begin{align}
                \mathcal{L}_n^{\dagger}=(-1)^{n+1}\overline{\mathcal{L}}_n,\:\:\overline{\mathcal{L}}_n^{\dagger}=(-1)^{n+1}\mathcal{L}_n\quad(n\in\mathbb{Z}), \label{eq: dS conjugation rule}
            \end{align}
            for which the unitarity is violated. Note that the Hamiltonian $H_{\rm dS}=\mathcal{L}_0+\overline{\mathcal{L}}_0$ is now anti-Hermitian $H^\dagger_{\rm dS}=-H_{\rm dS}$.
            Eq.\,\eqref{eq: dS conjugation rule} is compatible with the property $\overline{c}_{\text{dS}_3}=-c_{\text{dS}_3}^*$ of the Virasoro algebra \eqref{eq: Virasoro commutation relations in dS}. If we assume the non-chiral case $\overline{c}_{\text{dS}_3}=c_{\text{dS}_3}$, then we find that the central charge should be pure imaginary which agrees with the known fact in the dS$_3/$CFT$_2$ \cite{Cotler:2019nbi,Goodhew:2024eup}.
            We then rearrange the generators $\mathcal{L}_n$, $\overline{\mathcal{L}}_n$ ($n\in\mathbb{Z}$) to reproduce the BMS$_3$ algebra in the flat limit $l\to\infty$.
            The new generators are\footnote{
                Again, as in the AdS case, if we assume that the correct linear combinations that reproduce eqs.\,\eqref{eq: BMS-like commutators before the flat limit in dS} \eqref{eq: BMS-like commutators before the flat limit in dS}, $L_n(l)^{\dagger}=L_{-n}(l)$, and $M_n(l)^{\dagger}=M_{-n}(l)$ are of the form
                \begin{align}
                    \begin{aligned}
                        L_n(l) & =a_n\mathcal{L}_n+b_n\mathcal{L}_{-n}+c_n\overline{\mathcal{L}}_n+d_n\overline{\mathcal{L}}_{-n} & (a_n,b_n,c_n,d_n\in\mathbb{C}), \\
                        M_n(l) & =\alpha_n\mathcal{L}_n+\beta_n\mathcal{L}_{-n}+\gamma_n\overline{\mathcal{L}}_n+\delta_n\overline{\mathcal{L}}_{-n} & (\alpha_n,\gamma_n,\delta_n,\delta_n\in\mathbb{C}),
                    \end{aligned}
                \end{align}
                then there are infinitely many choices that give the solutions to the required conditions:
                \begin{gather}
                    L_n=\zeta^n\mathcal{L}_n-(-\zeta^*)^{-n}\overline{\mathcal{L}}_{-n},\;\;M_n=\frac{1}{l}i\Big(\zeta^n\mathcal{L}_n+(-\zeta^*)^{-n}\overline{\mathcal{L}}_{-n}\Big)\quad(\zeta\in\mathbb{C}^{\times}) \\
                    \text{or} \notag \\
                    L_n=-\eta^{-n}\mathcal{L}_{-n}+(-\eta^*)^n\overline{\mathcal{L}}_{n} ,\;\;M_n=-\frac{1}{l}i\Big(\eta^{-n}\mathcal{L}_{-n}+(-\eta^*)^n\overline{\mathcal{L}}_n\Big)\quad(\eta\in\mathbb{C}^{\times})
                \end{gather}
                However, we can choose and fix $\eta$ or $\zeta$ by demanding that the dual bulk isometries also matches to \eqref{eq: flat3killing} in the flat limit (See Appendix \ref{append: dS generators} for explicit form of isometries in dS$_3$).
                The answer is $\zeta=1$ in our convention.
            }
            \begin{align}
                 L_n(l)=\mathcal{L}_{-n}-(-1)^n\overline{\mathcal{L}}_{n} ,\;\;M_n(l)=\frac{1}{l}i\big(\mathcal{L}_{-n}+(-1)^n\overline{\mathcal{L}}_n\big),   \label{eq: generator relation between flat and dS}
            \end{align}
            whose commutators are the BMS-like
            \begin{align}
                \begin{aligned}
                    [L_m(l),L_n(l)] & =(m-n)L_{m+n}(l)+\frac{c_L}{12}(m^3-m)\delta_{m+n,0}, \\
                    [L_m(l),M_n(l)] & =(m-n)M_{m+n}(l)+\frac{c_M}{12}(m^3-m)\delta_{m+n,0}, \\
                    [M_m(l),M_n(l)] & =-\frac{1}{l^2}\Big((m-n)L_{m+n}(l)+\frac{c_L}{12}(m^3-m)\delta_{m+n,0}\Big).
                \end{aligned} \label{eq: BMS-like commutators before the flat limit in dS}
            \end{align}
            The new ``central charges" $c_L$, $c_R$ are given by
            \begin{align}
                c_L:=c_{\text{dS}_3}-\overline{c}_{\text{dS}_3}=0,\quad c_M:=-i\frac{c_{\text{dS}_3}+\overline{c}_{\text{dS}_3}}{l}=\frac{3}{G}.  \label{eq: BMS-like central charge from dS}
            \end{align}
            We also see from eq.\,\eqref{eq: dS conjugation rule} that eq.\,\eqref{eq: generator relation between flat and dS} properly satisfies the conjugation rule $L_n(l)^{\dagger}=L_{-n}(l)$, $M_n(l)^{\dagger}=M_{-n}(l)$ in BMS$_3$, and by taking $l\to\infty$, the commutators \eqref{eq: BMS-like commutators before the flat limit in dS} come down to the BMS$_3$ algebra \eqref{eq: bms_3 algebra}.
            \par We next verify that the highest weight representation in dS$_3$/CFT$_2$ also converts into the induced representation in the flat limit.
            One readily observes this when we rewrite the highest weight conditions \eqref{eq: highest weight condition in dS} in terms of the new generators \eqref{eq: generator relation between flat and dS}
            \begin{align}
                \begin{gathered}
                    L_0(l)|h,\overline{h}\rangle=(h-\overline{h})|h,\overline{h}\rangle,\:\:M_0(l)|h,\overline{h}\rangle=i\frac{h+\overline{h}}{l}|h,\overline{h}\rangle, \\
                    \Big(M_n(l)+\frac{i}{l}L_n(l)\Big)|h,\overline{h}\rangle=0,\:\:\Big(M_{-n}(l)-\frac{i}{l}L_{-n}(l)\Big)|h,\overline{h}\rangle=0\quad(n>0).
                \end{gathered}
            \end{align}
            Using the variables $\displaystyle\xi:=\lim_{l\to\infty}\frac{i}{l}(h+\overline{h}),\;\;\Delta:=\lim_{l\to\infty}(h-\overline{h})$,
            these conditions result in the induced condition
            \begin{align}
                M_0|\Delta,\xi\rangle=\xi|\Delta,\xi\rangle,\:M_n|\Delta,\xi\rangle=0\,(n\neq 0),\:L_0|\Delta,\xi\rangle=\Delta|\Delta,\xi\rangle,  \label{eq: BMS induced from dS highest}
            \end{align}
            where $\displaystyle L_n:=\lim_{l\to\infty}L_n(l),\;M_n:=\lim_{l\to\infty}M_n(l),\;|\Delta,\xi\rangle:=\lim_{l\to\infty}|h,\overline{h}\rangle$.
            As we focus on the scalar field $h=\overline{h}=\frac{1}{2}\pm\frac{1}{2}i\sqrt{m^2l^2-1}$ in this paper, the limit values $\xi=\pm m, \Delta=0$ are convergent as well.
            In this case, eq.\,\eqref{eq: BMS induced from dS highest} is identical to the scalar BMS induced condition \eqref{eq: bms induced rep} with eq.\,\eqref{eq: xi and weight for scalar in bms}.

        \subsection{Bulk local state correspondence}\label{subsec: states_correspondence_dS}
            In dS$_3$/CFT$_2$, the CFT state $\ket*{\Psi^{\Delta_\pm}_{\mathrm{dS,W}}(t=0,r=0)}$
            \footnote{Note that here we distinguish the scaling dimensions of the CFT$_2$ dual to dS$_3$, $\Delta_\pm=h+\bar{h}=1\pm i\sqrt{m^2l^2-1}$ from the conformal weight $\Delta$ in the CCFT$_2$.}
            dual to the bulk local scalar state at the origin $(t,\theta)=(0,0)$ must fulfill the constraint
            \begin{align}
                (\mathcal{L}_0-\overline{\mathcal{L}}_0)\ket*{\Psi^{\Delta_\pm}_{\mathrm{dS,W}}(t=0,r=0)}=0,\;\;(\mathcal{L}_{\pm 1}+\overline{\mathcal{L}}_{\mp 1})\ket*{\Psi^{\Delta_\pm}_{\mathrm{dS,W}}(t=0,r=0)}=0,  \label{eq: the constraint for the excitation at origin in dS}
            \end{align}
            which stand for the rotational and boost invariance of the local scalar excitation $\Psi^{\Delta_\pm}_{\mathrm{dS,W}}(t=0,r=0)$ in the bulk (see Appendix \ref{append: dS generators} for the explicit form of dS$_3$ isometries).
            The solution is given by \cite{Doi:2024nty}
            \begin{align}
                \ket*{\Psi_{\text{dS,W}}(t=0,r=0)}=e^{\frac{i \pi}{2}(\mathcal{L}_0+\overline{\mathcal{L}}_0-2h)}|I_h\rangle=\sum_{k=0}^{\infty}\prod_{j=1}^k\frac{e^{i\pi k}}{j\big(2h-1+j\big)}\mathcal{L}_{-1}^k\overline{\mathcal{L}}_{-1}^k|h,h\rangle,
            \end{align}
            where $|I_h\rangle$ is the Ishibashi state \eqref{eq: AdS  Ishibashi state}.
            Any other states located at the generic point $(t,r,\phi)$ are
            \begin{align}
                \ket*{\Psi_{\text{dS,W}}(t,r,\phi)}=e^{(\mathcal{L}_0+\overline{\mathcal{L}}_0)\frac{t}{l}}e^{\frac{i}{2}(\mathcal{L}_1-\overline{\mathcal{L}}_1-\mathcal{L}_{-1}+\overline{\mathcal{L}}_{-1})\operatorname{arcsinh}\frac{r}{l}}e^{i(\mathcal{L}_0-\overline{\mathcal{L}}_0)\phi}\ket*{\Psi_{\text{dS,W}}(t=0,r=0)}.
            \end{align}
            The next task is to explore the coordinates correspondence between dS$_3$ and the flat$_3$, along with the relation between dS$_3$ constraints and the flat$_3$ ones.

            \subsubsection*{(1) dS$_3$ origin $(t,r)=(0,0)$ $\to$ Flat$_3$ origin $(t,r)=(0,0)$}
            Firstly, the constraint for the excitation at origin in dS$_3$ is given by eq.\,\eqref{eq: the constraint for the excitation at origin in dS}.
            If we assume the existence of a putative limit state $\displaystyle\ket*{\phi(t,r,\phi)}:=\lim_{ l\to\infty}\ket*{\Psi_{\text{AdS}}(t,r,\phi)}$ (which will be shown later), this leads to the constraint
            \begin{align}
                L_0\ket*{\phi(0,0,0)}=0,\;\;L_{\pm 1}\ket*{\phi(0,0,0)}=0,
            \end{align}
            which represents the rotational invariance at origin \eqref{eq: origin equation} in the flat$_3$.

            \subsubsection*{(2) dS$_3$ $(t,r,\phi)=(t,0,0)$ $\to$ Flat$_3$ $(t,r,\phi)=(t,0,0)$}
            We also discuss the constraint that $\ket*{\Psi^{\Delta_\pm}_{\mathrm{dS,W}}(t,0,0)}$ must follow to avoid the divergences at the origin.
            The time translation operator is $e^{\frac{t}{l}(\mathcal{L}_0+\overline{\mathcal{L}}_0)}$, thus the constraint at the origin \eqref{eq: the constraint for the excitation at origin in dS} is transformed into
            \begin{align}
                (\mathcal{L}_0-\overline{\mathcal{L}}_0)\ket*{\Psi^{\Delta_\pm}_{\mathrm{dS,W}}(t,0,0)}=0,\;\;(e^{\pm \frac{t}{l}} \mathcal{L}_{\pm 1}+e^{\mp \frac{t}{l}}\overline{\mathcal{L}}_{\mp 1})\ket*{\Psi^{\Delta_\pm}_{\mathrm{dS,W}}(t,0,0)}=0. \label{eq: dS constraint at (t,0,0)}
            \end{align}
            Here, we use the identity $e^{-a\mathcal{L}_0}\mathcal{L}_{\pm 1} e^{a\mathcal{L}_0}=e^{\pm a}\mathcal{L}_{\pm 1}\;\;(a\in\mathbb{R})$.
            Thus, the flat limit of these conditions are given by
            \begin{align}   \label{eq:UR_limit_of_constraint}
                 L_0\ket*{\phi(t,0,0)}=0,\;\;(L_{\pm 1}\mp it M_{\pm 1})\ket*{\phi(t,0,0)}=0,
            \end{align}
            The second equations result from eq.\,\eqref{eq: generator relation between flat and dS}
            \begin{align}
                \lim_{l\to\infty}(e^{\pm \frac{t}{l}} \mathcal{L}_{\pm 1}+e^{\mp \frac{t}{l}}\overline{\mathcal{L}}_{\mp 1}) & =
                \lim_{l\to\infty}\left[\left(1\pm  \frac{t}{l}\right)\mathcal{L}_{\pm 1}+\left(1\mp \frac{t}{l}\right)\overline{\mathcal{L}}_{\mp 1}+O\left(\frac{1}{l^2}\right)\right] \nonumber \\[1ex]
                & =L_{\pm 1}\mp it M_{\pm 1}.
            \end{align}
            Since \eqref{eq: t equation} and \eqref{eq:UR_limit_of_constraint} are identical, the dS$_3$ constraints and  the flat$_3$ ones are equivalent via the flat limit. Particularly, we expect that the the bulk local states $\ket*{\Psi^{\Delta_\pm}_{\mathrm{dS,W}}(t,0,0)}$ and $\ket*{\Psi_\text{Flat}(t,0,0)}$ are related via the flat limit, more precisely given by $\displaystyle \ket*{\phi(t,0,0)}=\lim_{l\to\infty}\ket*{\Psi^{\Delta_\pm}_{\mathrm{dS,W}}(t,0,0)}=\ket*{\Psi_\text{Flat}(t,0,0)}$. \\
            \par The solution to eq.\,\eqref{eq: dS constraint at (t,0,0)} is
            \begin{align}
                \ket*{\Psi^{\Delta_\pm}_{\mathrm{dS,W}}(t,0,0)} & =e^{\frac{t}{l}(\mathcal{L}_0+\overline{\mathcal{L}}_0-2h)}\ket*{\Psi^{\Delta_\pm}_{\mathrm{dS,W}}(0,0,0)} =\sum_{k=0}^\infty \frac{e^{k(i\pi+2\frac{t}{l})}}{\Gamma (k+1) (2h)_k}\mathcal{L}_{-1}^k\bar{\mathcal{L}}_{-1}^k|h,h\rangle,
                \label{eq: dS bulk local state at (t,0,0)}
            \end{align}
            where $|I_h\rangle$ is the Ishibashi state \eqref{eq: AdS  Ishibashi state}.
            By analogy with eq.\,\eqref{eq: AdS descendants and fla descendants}, one can derive the following relation that relates AdS$_3$/CFT$_2$ descendants and the ``flat ones",
            \begin{align}
                \mathcal{L}_{-1}^k\overline{\mathcal{L}}_{-1}^k|h,h\rangle=\sum_{a=0}^k g_{k,a}\big(L_{-1}(l)\big)^a\big(L_{1}(l)\big)^a|h,h\rangle, \label{eq: dS descendants and fla descendants}
            \end{align}
            where the coefficient is given by
            \begin{equation}
               g_{k,a}= \frac{(-1)^{k-a} \Gamma (k+1)^2 \Gamma (k+2h)}{\Gamma (a+1)^2 \Gamma (-a+k+1) \Gamma (a+2h)}.
            \end{equation}
            This seems identical to the coefficient \eqref{eq: AdS flat descendant coefficient}, although the conformal weight is different: the AdS$_3$/CFT$_2$ conformal weight is $h=\frac{1}{2}+\frac{1}{2}\sqrt{m^2l^2+1}$, while the dS$_3$/CFT$_2$ conformal weight is given by $h=\frac{1}{2}\pm \frac{1}{2}i\sqrt{m^2l^2-1}$.
            The dS bulk local state at $(t,0,0)$ \eqref{eq: dS bulk local state at (t,0,0)} can be expanded in terms of the flat descendants
            \begin{align}
                \ket*{\Psi^{\Delta_\pm}_{\mathrm{dS,W}}(t,0,0)}=\sum_{a=0}^\infty\underbrace{\sum_{k=0}^\infty\frac{e^{k(i\pi+2\frac{t}{l})}}{\Gamma (k+1) (\Delta)_k}g_{k,a}}_{=:\,g_a}\big(L_{-1}(l)\big)^{a}\big(L_{1}(l)\big)^{a}|h,h\rangle.
            \end{align}
            To see if this state asymptotes to $|\Psi_{\text{Flat}}(t,0,0)\rangle$ as $l\to\infty$, we need to see the coefficient
            \begin{align}
                g_a=\sum_{k=0}^\infty \frac{e^{k(i\pi+2\frac{t}{l})}}{\Gamma (k+1) (2h)_k}g_{k,a}=\frac{(-1)^{-a} \, _2\tilde{F}_1\big(1,1;1-a;e^{\frac{2t}{l}}\big)}{\Gamma (a+1)^2 (2h)_a},
            \end{align}
            where $_2\tilde{F}_1(p,q;r;z)$ is the regularized hypergeometric function \eqref{eq: regularized hypergeometric function}.
            As $l\to\infty$,
            \begin{equation}
                \lim_{l\to\infty}\frac{g_a}{l}=-\frac{2^{-a-1}\xi(i \xi  t)^{-a-1}}{\Gamma (a+1)},\;\;\lim_{l\to\infty}\big(L_{-1}(l)\big)^{a}\big(L_{1}(l)\big)^{a}|h,h\rangle=L_{-1}^{a}L_{1}^{a}|0,\xi\rangle.
            \end{equation}
            Therefore, we conclude that
            \begin{equation}
            -\lim_{l\to\infty}\frac{2}{l}\ket*{\Psi^{\Delta_\pm}_{\mathrm{dS,W}}(t,0,0)}=\sum_{a=0}^\infty\frac{2^{-a}  \xi  (i \xi  t)^{-a-1}}{\Gamma (a+1)}L_{-1}^{a}L_{1}^{a}|0,\xi\rangle=\ket*{\Psi_\text{Flat}(t,0,0)}.
            \end{equation}
            At $t=0$, it is divergent when taking the limit because $ \, _2\tilde{F}_1(1,1;1-a,x)$ has the singularity at $x=1$.

\subsection{Green's functions correspondence}\label{greens_function_dS}
The geodesic length between two points in dS$_3$ is given by:
\begin{align}
    \cos\frac{D_{\text{dS}}(x_1,x_2)}{l_{\text{dS}}}=\cosh(\tau_1-\tau_2)\cos\theta_1\cos\theta_2+\cos(\phi_1-\phi_2)\sin\theta_1\sin\theta_2,
\end{align}
 where the metric reads
 \begin{equation}
     ds^2=-\cos^2\theta d\tau^2+d\theta^2+\sin^2 d\phi^2.
 \end{equation}
We define the Green's function obtained by the Wick rotation from  $G_\text{AdS}(x_1,x_2)$ as $G_{\text{dS, W}}^{\Delta_\pm}(x_1,x_2)$ \footnote{
We note that the bra state $\bra*{\Psi^{\Delta_\pm}_{\mathrm{dS,W}}(x_1)}$ is not $\big(\ket*{\Psi^{\Delta_\pm}_{\mathrm{dS,W}}(x_1)}\big)^\dagger$ but $\big(\ket*{\widehat{\Psi^{\Delta_\pm}_{\mathrm{dS,W}}}(x_1)}\big)^\dagger$ defined as dual state in \cite{Doi:2024nty}.
} :
\begin{align}
    G_{\text{dS,W}}^{\Delta_\pm}(x,x')=\ev*{\Psi^{\Delta_\pm}_{\mathrm{dS,W}}(x)|\Psi^{\Delta_\pm}_{\mathrm{dS,W}}(x')}=\frac{e^{\pm \mu D_{\text{dS}}/l_{\text{dS}}}}{2i \sin{D_{\text{dS}}/l_{\text{dS}}}}.
\end{align}
The flat limit of these Wick rotated Green's functions takes the form:
\begin{align}
    \lim_{l\to\infty}\frac{G_\text{dS,W}^{\Delta_{\pm}}(x,x')}{l_{\text{dS}}}=\lim_{l\to\infty}\frac{e^{\pm \mu D_{\text{dS}}(x,x')/l_{\text{dS}}}}{2il_{\text{dS}}\sin{D_{\text{dS}}(x,x')/l_{\text{dS}}}}=\frac{e^{\pm m D_{\text{flat}}}}{2iD_{\text{flat}}}.
\end{align}

On the other hand, the physical Green's function in the Bunch-Davies vacuum $G_\text{dS,E}(x,x')$ is given by a linear combination of $G_{\text{dS,W}}^{\Delta_\pm}$
\begin{align}
    G_\text{dS,E}(x,x')&=\frac{i}{4\pi \sinh{\pi\mu}}\left[G_{\text{dS,W}}^{\Delta_+}(x_A,x')-G_{\text{dS,W}}^{\Delta_-}(x,x_A')\right]\\
    &=\frac{\sinh{\mu(\pi-D_{\text{dS}}/l_{\text{dS}}})}{4\pi \sinh{\pi\mu}\sin{D_{\text{dS}}(x,x')/l_{\text{dS}}}}, \label{eq: Bunch-Davies vacuum correlator}
\end{align}
which is interpreted as the inner product of CPT invariant states in \cite{Doi:2024nty}:
\begin{align}
    |\Psi_{\text{E}}(x)\rangle=\frac{1}{2\sqrt{\pi\sinh\pi\mu}}\left[\frac{1}{\sqrt{i}}\ket*{\Psi^{\Delta_+}_{\text{dS,W}}(x)}
+\sqrt{i}\ket*{\Psi^{\Delta_{-}}_{\text{dS,W}}(x_A)}\right].  \label{psiEa}
\end{align}
Indeed, the inner product of $|\Psi_{\text{E}}(x)\rangle$ gives the physical Green's function as follows:
\begin{align}
     \langle \Psi_{\rm E}(x)|\Psi_{\rm E}(x')\rangle&=\frac{i}{4\pi\sinh\pi\mu}\left[\ev*{\widehat{\Psi_{\text{dS,W}}^{\Delta_{+}}}(x)|\Psi^{\Delta_{-}}_{\text{dS,W}}(x'_A)}-\ev*{\widehat{\Psi_{\text{dS,W}}^{\Delta_{-}}}(x_A)|\Psi^{\Delta_{+}}_{\text{dS,W}}(x')}\right] \notag\\
&= \frac{i}{4\pi\sinh\pi\mu}\left[\frac{e^{\mu\left(\pi-D_{\rm dS}(x,x')\right)}}{2i\sin D_{\rm dS}(x,x')}-\frac{e^{-\mu\left(\pi-D_{\rm dS}(x,x')\right)}}{2i\sin D_{\rm dS}(x,x')}\right]\notag\\
&=\frac{\sinh\mu\left(\pi-D_{\rm dS}(x,x')\right)}{4\pi \sinh \mu\pi \sin D_{\rm dS}(x,x')}.\label{GEds}
\end{align}
In the flat limit, this physical Green’s function reduces to:
\begin{align}
    \lim_{l\to\infty}\frac{G_\text{dS,E}(x,x')}{l_{\text{dS}}} & =\lim_{l\to\infty}\frac{\sinh{\pi\mu}\cosh{\mu D_{\text{dS}}/l_{\text{dS}}}-\sinh{\mu D_{\text{dS}}/l_{\text{dS}}}\cosh{\pi\mu}}{4\pi l_{\text{dS}}\sinh{\pi\mu}\sin{D_{\text{dS}}(x,x')/l_{\text{dS}}}} \nonumber \\
        & = \frac{\cosh{m D_{\text{flat}}}-\sinh{m D_{\text{flat}}}}{4\pi D_{\text{flat}}} \nonumber \\
        & = \frac{e^{-m D_{\text{flat}}}}{4\pi D_{\text{flat}}}.
\end{align}
Therefore, the Green’s function on the flat$_3$ spacetime is correctly obtained by taking the flat limit of the inner product of the CPT-invariant states.
\footnote{We note, however, that here as well we have not determined the explicit flat limit solution for the bra state and thus cannot compute the inner product directly; see also the discussion in Section \ref{section: conclusion and outlook}.}

\section{Conclusions and discussions}
\label{section: conclusion and outlook}

As an approach to flat holography, Carrollian CFTs (CCFTs) are proposed as the potential dual quantum field theory in a spacetime with one dimension lower than the  bulk flat space. This paper advances this concept by focusing on the bulk reconstruction of massive scalar excitations, particularly in the Flat$_3$/CCFT$_2$ correspondence. The method involves expressing the bulk local states using CCFT states and listing the equations derived from rotational invariance constraints. Two representations are discussed in CCFTs: the highest weight representation and the induced representation. To determine the appropriate representation for the flat holography, we solve the constraint equations for the bulk scalar excitation in both representations, concluding that the induced representation provides the correct solution. The unitarity of the induced representation suggests that the gravitational theory in the flat$_3$ spacetime is unitary, consistent with the absence of gravitational radiation leakage from null infinity, a feature unique to the three dimensions.

To delve deeper, we need to examine the properties of the bulk local states we derived. The first challenge we encounter is that the inner product in the induced module cannot be calculated as it is in the highest weight module. To address this, we introduce the dual basis and identify the state $^\vee\langle 0|$ as the bra state for the bulk scalar excitation at the origin. Using this dual state, we can calculate the inner product of the two bulk local states at different positions, which aligns well with the bulk scalar two-point function. This, in turn, determines the bulk flat metric using the information metric of the bulk local states. Additionally, the spectrum of the scalar excitation in the bulk matches the dictionary:
\begin{equation}
\xi = m, \ \Delta = 0,
\end{equation} where $\xi$ and $\Delta$ are the boost charge and scaling dimension of the induced module in CCFT$_2$, and $m$ is the mass parameter for the scalar in the bulk.

The reconstruction of local states in AdS/CFT and dS/CFT correspondences have been discussed in the previous literatures \cite{Miyaji:2015fia,Doi:2024nty}. We explore whether a valid flat limit exists from the (A)dS$_3$/CFT$_2$ correspondence, leading to the Flat$_3$/CCFT$_2$ correspondence. By comparing the Killing vectors in (A)dS$_3$ in the large (A)dS radius limit and the flat$_3$ spacetime, we establish the relationship between CFT generators and CCFT generators. This flat limit appears to be novel. Based on this limit, we verify that the bulk local states in the flat$_3$ spacetime can be derived from the ones in the (A)dS$_3$ spacetime, as well as the bulk two-point function. We hope this new flat limit will inspire further research on other various aspects in the flat holography.

The discussions and future directions are listed as follows.

\subsection*{Flat limit of dual basis}
In Sections \ref{sec:flat_limit_AdS} and \ref{sec:flat_limit_dS}, we showed that the kets and Green's functions for local scalar excitations on the flat$_3$ spacetime can be obtained exactly as the flat limit of their (A)dS$_3$ counterparts.
However, we note that this limit does not establish a full one-to-one correspondence between the local excitation states in (A)dS$_3$ and those in Flat$_3$.
In particular, we have no solutions for the flat limit of the bra states.
This is probably related to the following superficially inconsistent problems, and resolving this puzzle is our immediate priority.

In section \ref{flat_ccft_exitation}, to compute the inner product with the ket states, we introduced the somewhat tricky dual basis $^{\vee}\bra*{0}$, which ultimately yields $\bra*{\phi(0,0,0)}$.
As has been emphasized, it is difficult to obtain $\ket*{\phi(0,0,0)}$ or $\bra*{\phi(0,0,0)}$ directly by taking the flat limit from (A)dS$_3$, and so far we have not succeeded in deriving the bra states in this way.
The failure of the naive flat limit also manifests itself in the required scaling. As seen in Sections \ref{greens_function_AdS} and \ref{subsec: states_correspondence_AdS}, both the ket states and Green’s functions must scale as $1/l$ in the flat limit.
This suggests that no analogous scaling occurs for the bra states, and hence that their flat‐limit structure differs from that of the kets.
Similarly, the same issue arises in the dS$_3$ case.

Moreover, in the dS$_3$ the structure of the dual space is even more intricate.
As noted in \cite{Doi:2024nty}, there are four ket states,
$
\ket*{\Psi^{\Delta_\pm}_{\mathrm{dS,W}}(x)}
\quad\text{and}\quad
\ket*{\widehat{\Psi^{\Delta_\pm}_{\mathrm{dS,W}}}(x)},
$
whose conjugates
$\bra*{\widehat{\Psi^{\Delta_\pm}_{\mathrm{dS,W}}}(x)}
:=(\ket*{\Psi^{\Delta_\pm}_{\mathrm{dS,W}}(x)})^\dagger,
\quad
\bra*{\Psi^{\Delta_\pm}_{\mathrm{dS,W}}(x)}
:=(\ket*{\widehat{\Psi^{\Delta_\pm}_{\mathrm{dS,W}}}(x)})^\dagger
$
obey the complex conjugation relations shown in Table \ref{tab:conj_relations}.
\cite{Doi:2024nty} conjectured that the physical Hilbert space is obtained by projecting onto the CPT‐invariant states; although one would naturally expect the flat limit of this projected space to yield the flat‐holography Hilbert space, the issue may be yet more involved.

In summary, our immediate challenge is to derive the bra states that successfully resolve these obstacles via the flat limit and thereby clarify the correspondence between the two Hilbert spaces.
\footnote{We originally assumed that under the antipodal map the geodesic‐distance dependence in the Green’s function transforms as $D(x,x')\to\pi-D(x,x')$, since geometrically $D(x_A,x')=\pi-D(x,x')$.
This follows the prescription of \cite{Doi:2024nty}.

However, a careful examination of the analyticity of the Wick rotated Green’s function may suggest that one could instead have $D(x,x')\to\pi+D(x,x')$.
In that case, Table \ref{tab:conj_relations} must be modified as below one.
Although this alternative looks more natural than the sign-alternating pattern in Table \ref{tab:conj_relations}, the correct choice remains unknown.
Whether one can derive a flat limit for the dual states that appears most natural may thus serve as a criterion for fixing the dual state construction in dS.

\begin{align*}
    \begin{array}{|c|c|c|c|c|}
    \hline
       & \ket*{\Psi^{\Delta_+}_{\mathrm{dS,W}}(x)}
       & \ket*{\Psi^{\Delta_-}_{\mathrm{dS,W}}(x)}
       & \ket*{\widehat{\Psi^{\Delta_-}_{\mathrm{dS,W}}}(x)}
       & \ket*{\widehat{\Psi^{\Delta_+}_{\mathrm{dS,W}}}(x)} \\
    \hline
    \bra*{\Psi^{\Delta_+}_{\mathrm{dS,W}}(x)}
      & G_\text{dS,W}^{\Delta_{+}}(x,x')
      & 0
      & G_\text{dS,W}^{\Delta_{+}}(x,x')
      & 0 \\
    \hline
    \bra*{\Psi^{\Delta_-}_{\mathrm{dS,W}}(x)}
      & 0
      & G_\text{dS,W}^{\Delta_{-}}(x,x')
      & 0
      & G_\text{dS,W}^{\Delta_{-}}(x,x') \\
    \hline
    \bra*{\widehat{\Psi^{\Delta_-}_{\mathrm{dS,W}}}(x)}
      & G_\text{dS,W}^{\Delta_{+}}(x,x')
      & 0
      & G_\text{dS,W}^{\Delta_{+}}(x,x')
      & 0 \\
    \hline
    \bra*{\widehat{\Psi^{\Delta_+}_{\mathrm{dS,W}}}(x)}
      & 0
      & G_\text{dS,W}^{\Delta_{-}}(x,x')
      & 0
      & G_\text{dS,W}^{\Delta_{-}}(x,x') \\
    \hline
    \end{array}
\end{align*}

}

\begin{table}[ht]
  \centering
  \renewcommand{\arraystretch}{1.2}
  \begin{tabular}{|c|c|c|c|c|}
    \hline
       & $\ket*{\Psi^{\Delta_+}_{\mathrm{dS,W}}(x)}$
       & $\ket*{\Psi^{\Delta_-}_{\mathrm{dS,W}}(x)}$
       & $\ket*{\widehat{\Psi^{\Delta_-}_{\mathrm{dS,W}}}(x)}$
       & $\ket*{\widehat{\Psi^{\Delta_+}_{\mathrm{dS,W}}}(x)}$ \\
    \hline
    $\bra*{\Psi^{\Delta_+}_{\mathrm{dS,W}}(x)}$
      & $G_\text{dS,W}^{\Delta_{+}}(x,x')$
      & $0$
      & $G_\text{dS,W}^{\Delta_{-}}(x,x')$
      & $0$ \\
    \hline
    $\bra*{\Psi^{\Delta_-}_{\mathrm{dS,W}}(x)}$
      & $0$
      & $G_\text{dS,W}^{\Delta_{-}}(x,x')$
      & $0$
      & $G_\text{dS,W}^{\Delta_{+}}(x,x')$ \\
    \hline
    $\bra*{\widehat{\Psi^{\Delta_-}_{\mathrm{dS,W}}}(x)}$
      & $G_\text{dS,W}^{\Delta_{-}}(x,x')$
      & $0$
      & $G_\text{dS,W}^{\Delta_{+}}(x,x')$
      & $0$ \\
    \hline
    $\bra*{\widehat{\Psi^{\Delta_+}_{\mathrm{dS,W}}}(x)}$
      & $0$
      & $G_\text{dS,W}^{\Delta_{+}}(x,x')$
      & $0$
      & $G_\text{dS,W}^{\Delta_{-}}(x,x')$ \\
    \hline
  \end{tabular}
    \caption{Inner products for the states dual to dS$_3$ scalar. Each inner product may carry an overall constant prefactor, which we have omitted.}
    \label{tab:conj_relations}
\end{table}

\subsection*{Operator reconstruction}
It has been demonstrated in \cite{Goto:2016wme} that in the AdS$_3$/CFT$_2$ context, the bulk reconstruction of the local state \cite{Miyaji:2015fia} is equivalent to the HKLL bulk reconstruction of operators \cite{Hamilton:2005ju,Hamilton:2006az}, with the state-operator correspondence playing a crucial role. A natural question arises: can a similar operator reconstruction be established in the Flat$_3$/CCFT$_2$ and linked to the bulk local state reconstruction? However, when discussing induced representations, they are often referred to as representations of states, and it remains unclear how to populate these representations with operators, as there are no established state-operator correspondences. Identifying operators to fill the induced representation is an intriguing challenge. Non-local operators in CCFTs are potential candidates. Additionally, exploring an HKLL-like reconstruction in the  Flat$_3$/CCFT$_2$ using mode sum or Green's function methods is also of interest. These topics are left for future research.

\subsection*{Bulk reconstruction in higher dimensional spacetime}
In this paper, we focus on the discussion in the Flat$_3$/CCFT$_2$, which simplifies the analysis due to the relatively simple algebra. Additionally, discussing the three-dimensional bulk avoids the issue of gravitational radiation leaking from the null boundary, a problem that can occur in higher-dimensional spacetimes. This potentially introduces external source and leads to the non-unitarity in the dual field theory. To advance, it is natural to consider a similar setup and the reconstruction of the bulk local state in higher-dimensional spacetimes. We find that in the flat$_4$ spacetime, the conjugation relations of the generators are unusual, leading to non-unitarity. On the other hand, we can propose a similar induced representation in CCFT$_3$ and examine the constraint equations. The flat limit is also expected to be helpful. This direction is ongoing.

\acknowledgments
We are grateful to Song He, Wenxin Lai, Weibo Mao and Wei Song for valuable discussions.
This work is supported by MEXT KAKENHI Grant-in-Aid for Transformative Research Areas (A) through the ``Extreme Universe'' collaboration: Grant Number 21H05187. TT is also supported by JSPS Grant-in-Aid for Scientific Research (B) No.~25K01000. YS is supported by Grant-in-Aid for JSPS Fellows No.23KJ1337.
KS is supported by Grant-in-Aid for JSPS Fellows No.25KJ1498.

\appendix

\section{Useful formulas in Flat$_3$/CCFT$_2$}
\label{append: formulas}
This appendix compiles the formulas commonly used in calculations throughout the main body of this paper. All formulas are derived directly from the BMS$_3$ algebra. We have omitted the commutators that are straightforwardly zero.
\begin{equation}
    [L_1,M_{-1}^k]=2kM_{-1}^{k-1}M_0.
\end{equation}
\begin{equation}
    [L_{-1},M_{1}^k]=-2kM_{1}^{k-1}M_0.
\end{equation}
\begin{equation}
    [L_0,M_{-1}^k]=kM_{-1}^k.
\end{equation}
\begin{equation}
    [L_0,L_{-1}^k]=kL_{-1}^k.
\end{equation}
\begin{equation}
    [L_0,M_{1}^k]=-kM_{1}^k.
\end{equation}
\begin{equation}
    [L_0,L_{1}^k]=-kL_{1}^k.
\end{equation}
\begin{equation}    \label{eq: L-1 and L1}
    [L_{-1},L_{1}^k]=k(k-1)L_{1}^{k-1}-2kL_{1}^{k-1}L_0.
\end{equation}
\begin{equation}
    [L_1,L_{-1}^k]=k(k-1)L_{-1}^{k-1}+2kL_{-1}^{k-1}L_0.
\end{equation}
\begin{equation}
    [M_1,L_{-1}^k]=k(k-1)L_{-1}^{k-2}M_{-1}+2kL_{-1}^{k-1}M_0.
\end{equation}
\begin{equation}    \label{eq: M-1 and L1}
    [M_{-1},L_{1}^k]=k(k-1)L_{1}^{k-2}M_{1}-2kL_{1}^{k-1}M_0.
\end{equation}
For a induced primary state $|\xi\rangle$ with zero $\Delta$, defining
\begin{equation}
|k\rangle=L_{-1}^kL_1^k|\xi\rangle,
\end{equation}
we have
\begin{equation}\label{eq: m0 action on k}
M_0|k\rangle=-2k^2\xi|k-1\rangle+\xi|k\rangle,
\end{equation}
\begin{equation}
    M_0^n|k\rangle=\sum_{i=0}^k\frac{(-2)^i\xi^n}{i!}(k-i+1)_i^2(n-i+1)_i|k-i\rangle,
\end{equation}
\begin{equation}\label{eq: eam0 action on k}
    e^{aM_0}|k\rangle=\sum_{i=0}^k \frac{(-2)^i e^{a \xi } (a \xi )^i
   \left((-i+k+1)_i\right){}^2}{i!}|k-i\rangle=\sum_{i=0}\frac{e^{a \xi } (-2)^{k-i}
   \left((i+1)_{k-i}\right){}^2 (a \xi
   )^{k-i}}{(k-i)!}|i\rangle.
\end{equation}
Defining
\begin{equation}
|i,j\rangle=L_{-1}^iL_1^j|\xi\rangle,
\end{equation}
we have
\begin{equation} \label{eq: m_1 action}
M_1|i,j\rangle=2\xi i(-(i-1)j|i-2,j-1\rangle+|i-1,j\rangle),
\end{equation}
\begin{equation}\nonumber
M_{-1}|i,j\rangle=-2\xi j|i,j-1\rangle,
\end{equation}
\begin{equation}\nonumber
    L_1|i,j\rangle=i(i-1-2j)|i-1,j\rangle+|i,j+1\rangle,
\end{equation}
\begin{equation}\label{eq: action of generators}
L_{-1}|i,j\rangle=|i+1,j\rangle.
\end{equation}
For the induced primary state $|\xi\rangle$ with vanishing $\Delta$, particular attention should be given to the following observation,
\begin{equation}
L_{-1}^kL_{1}^k|\xi\rangle=L_{1}^kL_{-1}^k|\xi\rangle
\end{equation}
which has been used frequently in the calculations.

        \subsection{Proof of the orthogonality relation} \label{subappend:orthogonality}
            We prove the orthogonality relation \eqref{eq: inner product} for $\xi\neq 0$.
            The normalization $\ ^\vee\!\langle 0|\xi\rangle=1$ fixes the remaining scale of the dual primary state.
            Its defining annihilation conditions then restrict the pairings to
            \begin{equation}\label{eq: dual primary pairing}
                \ ^\vee\!\langle 0|i,j\rangle=\delta_{i,0}\delta_{j,0}.
            \end{equation}
            The calculation begins with the pairing of $\ ^\vee\!\langle 0|M_1^rM_{-1}^r$ and the diagonal ket state $|k\rangle=|k,k\rangle$.
            By repeating the $M_{-1}$ action, we get
            \begin{equation}
                M_{-1}^r|k,k\rangle=(-2\xi)^r\frac{k!}{(k-r)!}|k,k-r\rangle,\qquad 0\leq r\leq k.
            \end{equation}
            In each of the subsequent $r$ actions of $M_1$, the two terms in its action in eq.\,\eqref{eq: m_1 action} lower $(i,j)$ by $(2,1)$ and $(1,0)$, respectively.
            A contribution to $|0,0\rangle$ must therefore select the first term $k-r$ times and the second term $2r-k$ times, which is possible only when $r\leq k\leq 2r$.
            The two lowering operations commute on $|i,j\rangle$.
            Each ordering with $k-r$ selections of the first term contributes $(-1)^{k-r}(2\xi)^r k!(k-r)!$, and the $\binom{r}{k-r}$ possible orderings sum to
            \begin{equation}
                \ ^\vee\!\langle 0|M_1^r|k,k-r\rangle=(-1)^{k-r}(2\xi)^r k!(k-r)!\binom{r}{k-r}.
            \end{equation}
            The resulting matrix element, including the preceding action of $M_{-1}^r$, is
            \begin{equation}\label{eq: triangular pairing}
                \ ^\vee\!\langle 0|M_1^rM_{-1}^r|k\rangle=(-1)^k(2\xi)^{2r}(k!)^2\binom{r}{k-r},
            \end{equation}
            where the binomial coefficient is understood to vanish unless $r\leq k\leq 2r$.
            In particular, only finitely many terms in the definition of $\ ^\vee\!\langle k'|$ act nontrivially on any fixed $|k\rangle$.

            For $k'=0$, $C_{0r}=\delta_{r,0}$ reduces eq.\,\eqref{eq: dual descendant states} to $\ ^\vee\!\langle 0|$, whose pairing in eq.\,\eqref{eq: dual primary pairing} is $\delta_{k,0}$.
            For $k'\geq1$, the definition \eqref{eq: dual descendant states} and the matrix element \eqref{eq: triangular pairing} express the inner product as
            \begin{equation}
                \ ^\vee\!\langle k'|k\rangle
                =(2\xi)^{2k'}(k!)^2\sum_{r=k'}^k(-1)^{k-r}C_{k'r}\binom{r}{k-r}.
            \end{equation}
            This sum is empty for $k<k'$ and equals $C_{k'k'}=1$ for $k=k'$.
            If $k>k'$, the coefficient in eq.\,\eqref{eq: Ckr coefficient} satisfies
            \begin{equation}\label{eq: finite difference coefficient}
                C_{k'r}\binom{r}{k-r}
                =\frac{k'}{(k-k')!}\binom{k-k'}{r-k'}
                \prod_{a=1}^{k-k'-1}(2r-k'-a).
            \end{equation}
            To show that $\ ^\vee\!\langle k'|k\rangle$ vanishes for $k>k'$, define the forward difference of a polynomial by
            \begin{equation}
                \Delta p(r):=p(r+1)-p(r).
            \end{equation}
            The leading powers in this subtraction cancel, so each application of $\Delta$ lowers the degree by one; hence $\Delta^n p=0$ whenever $\deg p<n$.
            The binomial expansion of the $n$ successive differences is
            \begin{equation}
                \Delta^n p(k')
                =\sum_{j=0}^n(-1)^{n-j}\binom{n}{j}p(k'+j).
            \end{equation}
            With $n=k-k'$, $j=r-k'$, and $p(r)$ equal to the product in eq.\,\eqref{eq: finite difference coefficient}, that substitution reduces the alternating sum in the inner product to $k'\Delta^n p(k')/n!$.
            Since $p(r)$ has degree $k-k'-1<n$, the sum vanishes.
            Consequently,
            \begin{equation}
                \ ^\vee\!\langle k'|k\rangle
                =(2\xi)^{2k'}(k!)^2\delta_{k,k'}
                =\bigl(k!(2\xi)^k\bigr)^2\delta_{k,k'},
            \end{equation}
            which is the desired equation \eqref{eq: inner product}.

            As a concrete example, the $k'=2$ specialization of eq.\,\eqref{eq: dual descendant states}, with the coefficients in eq.\,\eqref{eq: Ckr coefficient}, is
            \begin{equation}
                \ ^\vee\!\langle 2|
                =\ ^\vee\!\langle 0|\left[
                M_1^2M_{-1}^2-\frac{2}{(2\xi)^2}M_1^3M_{-1}^3
                +\frac{5}{(2\xi)^4}M_1^4M_{-1}^4-\cdots\right].
            \end{equation}
            By eq.\,\eqref{eq: triangular pairing}, its pairings with $|1\rangle$, $|2\rangle$, and $|3\rangle$ are
            \begin{equation}
                \begin{aligned}
                    \ ^\vee\!\langle 2|1\rangle&=0,\\
                    \ ^\vee\!\langle 2|2\rangle&=4(2\xi)^4=N_2,\\
                    \ ^\vee\!\langle 2|3\rangle&=-72(2\xi)^4+72(2\xi)^4=0.
                \end{aligned}
            \end{equation}
            Thus the first nontrivial off-diagonal pairing vanishes by cancellation between the $r=2$ and $r=3$ terms, rather than because each term vanishes separately.

\section{Technical details in (A)dS$_3$/CFT$_2$}
    \label{app: technical details}
        \subsection{AdS$_3$ bulk isometries} \label{append: AdS generators}
            For the global coordinate of AdS$_3$
            \begin{align}
                g_{\text{AdS}_3}=l^2(-\cosh^2\rho d\tau^2+d\rho^2+\sinh^2\rho d\phi^2),
            \end{align}
            the isometry generators read
            \begin{align}
                \begin{gathered}
                    \mathcal{L}_0=\frac{i}{2}(\partial_{\tau}+\partial_{\phi}),\:\:\overline{\mathcal{L}}_0=\frac{i}{2}(\partial_{\tau}-\partial_{\phi}), \\
                    \mathcal{L}_{\pm 1}=\frac{i}{2}e^{\pm i(\tau+\phi)}\Big[\frac{\sinh\rho}{\cosh\rho}\partial_{\tau}+\frac{\cosh\rho}{\sinh\rho}\partial_{\phi}\mp i\partial_{\rho}\Big], \\
                    \overline{\mathcal{L}}_{\pm 1}=\frac{i}{2}e^{\pm i(\tau-\phi)}\Big[\frac{\sinh\rho}{\cosh\rho}\partial_{\tau}-\frac{\cosh\rho}{\sinh\rho}\partial_{\phi}\mp i\partial_{\rho}\Big].
                \end{gathered}
            \end{align}
            After the coordinate transformation $t:=l\tau$, $r:=l\sinh\rho$,
            \begin{align}
                \partial_{\tau}=l\partial_t,\:\:\partial_{\rho}=\sqrt{l^2+r^2}\partial_r,
            \end{align}
            and the generators are expressed as
            \begin{align}
                \begin{gathered}
                    \mathcal{L}_0=\frac{i}{2}(l\partial_t+\partial_{\phi}),\:\:\overline{\mathcal{L}}_0=\frac{i}{2}(l\partial_t-\partial_{\phi}), \\
                    \mathcal{L}_{\pm 1}=\frac{i}{2}e^{\pm i(\frac{t}{l}+\phi)}\Big[\frac{r}{\sqrt{l^2+r^2}}\partial_t+\frac{\sqrt{l^2+r^2}}{r}\partial_{\phi}\mp i\sqrt{l^2+r^2}\partial_r\Big], \\
                    \overline{\mathcal{L}}_{\pm 1}=\frac{i}{2}e^{\pm i(\frac{t}{l}-\phi)}\Big[\frac{r}{\sqrt{l^2+r^2}}\partial_t-\frac{\sqrt{l^2+r^2}}{r}\partial_{\phi}\mp i\sqrt{l^2+r^2}\partial_r\Big]
                \end{gathered}
            \end{align}
            Further transformation $u:=t-l\arctan{\frac{r}{l}}$ leads to the metric
            \begin{align}
                g_{\text{AdS}_3}=-\Big(1+\frac{r^2}{l^2}\Big)du^2-2dudr+r^2d\phi^2, \label{eq: AdS metric in u,r,phi}
            \end{align}
            and the isometries are
            \begin{align}
                \begin{gathered}
                \mathcal{L}_0=\frac{i}{2}(l\partial_u+\partial_{\phi}),\:\:\overline{\mathcal{L}}_0=\frac{i}{2}(l\partial_u-\partial_{\phi}), \\
                \mathcal{L}_{\pm 1}=\frac{i}{2}e^{\pm i(\frac{u}{l}+\arctan\frac{r}{l}+\phi)}\sqrt{r^2+l^2}\Big[\frac{l}{r\mp il}\partial_t+\frac{1}{r}\partial_{\phi}\mp i\partial_r\Big], \\
                \overline{\mathcal{L}}_{\pm 1}=\frac{i}{2}e^{\pm i(\frac{u}{l}+\arctan\frac{r}{l}-\phi)}\sqrt{r^2+l^2}\Big[\frac{l}{r\mp il}\partial_t-\frac{1}{r}\partial_{\phi}\mp i\partial_r\Big].
            \end{gathered}
            \end{align}

        \subsection{dS$_3$ bulk isometries} \label{append: dS generators}
            Consider the static patch of the de Sitter space
              \begin{align}
                  g_{\text{dS}}=l^2\big(-\cos^2\theta~d\tau^2+d\theta^2+\sin^2\theta~d\phi^2\big).
              \end{align}
              In this coordinate, the bulk $SL(2,\mathbb{C})$ isometry generators are
              \begin{align}
                  \begin{gathered}
                      \mathcal{L}_0=\frac{1}{2}(\partial_{\tau}+i\partial_{\phi}),\:\:\overline{\mathcal{L}}_0=\frac{1}{2}(\partial_{\tau}-i\partial_{\phi}), \\
                      \mathcal{L}_{\pm 1}=\frac{i}{2}e^{\mp(\tau-i\phi)}\Big[\frac{\sin\theta}{\cos\theta}\partial_{\tau}-i\frac{\cos\theta}{\sin\theta}\partial_{\phi}\mp\partial_{\theta}\Big], \\
                      \overline{\mathcal{L}}_{\pm 1}=\frac{i}{2}e^{\mp(\tau+i\phi)}\Big[\frac{\sin\theta}{\cos\theta}\partial_{\tau}+i\frac{\cos\theta}{\sin\theta}\partial_{\phi}\mp\partial_{\theta}\Big]
                  \end{gathered}
              \end{align}
              Under the coordinate transformation $t:=l\tau$, $r:=l\sin\theta~(0\leq r\leq l)$, the static patch metric reads
              \begin{align}
                  g_{\text{dS}}=-\Big(1-\frac{r^2}{l^2}\Big)dt^2+\frac{1}{1-\frac{r^2}{l^2}}dr^2+r^2d\phi^2,
              \end{align}
              and the isometry generators are given by
              \begin{align}
                  \begin{gathered}
                      \mathcal{L}_0=\frac{1}{2}(l\partial_t+i\partial_{\phi}),\:\:\overline{\mathcal{L}}_0=\frac{1}{2}(l\partial_t-i\partial_{\phi}), \\
                      \mathcal{L}_{\pm 1}=\frac{i}{2}e^{\pm(-\frac{t}{l}+i\phi)}\Big[\frac{lr}{\sqrt{l^2-r^2}}\partial_t-i\frac{\sqrt{l^2-r^2}}{r}\partial_{\phi}\mp\sqrt{l^2-r^2}\partial_r\Big], \\
                      \overline{\mathcal{L}}_{\pm 1}=\frac{i}{2}e^{\pm(-\frac{t}{l}-i\phi)}\Big[\frac{lr}{\sqrt{l^2-r^2}}\partial_t+i\frac{\sqrt{l^2-r^2}}{r}\partial_{\phi}\mp\sqrt{l^2-r^2}\partial_r\Big].
                  \end{gathered}
              \end{align}
              Further coordinate changes $\displaystyle u:=t-l\operatorname{arctanh}\frac{r}{l}$ enable us to rewrite the metric in a similar way as the retarded coordinate of the 3d Minkowski metric \eqref{eq: 3d flat metric in the retarded coordinate}
              \begin{align} \label{eq: dS metric in u,r,phi}
                  g_{\text{dS}}=-\Big(1-\frac{r^2}{l^2}\Big)du^2-2dudr+r^2d\phi^2,
              \end{align}
              which precisely matches with it in the flat limit $l\to\infty$.
              Meanwhile the differential operators read
              \begin{align}
                  \partial_t\mapsto\partial_u,\:\:\partial_{r}\mapsto-\frac{l^2}{l^2-r^2}\partial_u+\partial_r,\:\:\partial_{\phi}\mapsto\partial_{\phi},
              \end{align}
              and the isometry generators are
              \begin{align}
                  \begin{gathered}
                      \mathcal{L}_0=\frac{1}{2}(l\partial_u+i\partial_{\phi}),\:\:\overline{\mathcal{L}}_0=\frac{1}{2}(l\partial_u-i\partial_{\phi}), \\
                      \mathcal{L}_{\pm 1}=\frac{i}{2}e^{\pm(-\frac{u}{l}-\operatorname{arctanh}\frac{r}{l}+i\phi)}\sqrt{l^2-r^2}\Big[\frac{l}{\pm l-r}\partial_u-i\frac{1}{r}\partial_{\phi}\mp\partial_r\Big], \\
                      \overline{\mathcal{L}}_{\pm 1}=\frac{i}{2}e^{\pm(-\frac{u}{l}-\operatorname{arctanh}\frac{r}{l}-i\phi)}\sqrt{l^2-r^2}\Big[\frac{l}{\pm l-r}\partial_u+i\frac{1}{r}\partial_{\phi}\mp\partial_r\Big]
                  \end{gathered}
              \end{align}
              The linear combinations \eqref{eq: generator relation between flat and dS} of these generators reproduce the flat BMS$_3$ generators \eqref{eq: flat3killing} in the flat limit. In computing this, one will frequently use the Taylor expansion $\operatorname{arctanh}x=x+\mathcal{O}(x^3)~(x\to0)$. \\

        \subsection{Proof of eq.\,\eqref{eq: AdS descendants and fla descendants}} \label{append: proof of descendant relation}
            In this appendix, we will prove the following relation between AdS$_3$/CFT$_2$ descendant and Flat$_3$/BMS$_2$ presented in eq.\,\eqref{eq: AdS descendants and fla descendants}
            \begin{align}\label{eq: relation des}
                \boxed{                       \mathcal{L}_{-1}^k\overline{\mathcal{L}}_{-1}^k|h,h\rangle=\sum_{a=0}^k \frac{(-1)^{k-a} \Gamma (k+1)^2 \Gamma (k+2h)}{\Gamma (a+1)^2 \Gamma (-a+k+1) \Gamma (a+2h)}\big(L_{-1}(l)\big)^a\big(L_{1}(l)\big)^a|h,h\rangle
                }.
            \end{align}
            To begin, we see from the relation \eqref{eq: generator relation between flat and AdS} and the highest weight condition \eqref{eq: highest weight condition in AdS represented in flat generators} that
            \begin{align}\nonumber\label{eq: relation k}
     \mathcal{L}_{-1}^k|h,h\rangle&=\frac{1}{2^k}\big(L_{-1}(l)+lM_{-1}(l)\big)^k|h,h\rangle=\big(L_{-1}(l)\big)^k|h,h\rangle,\\
 \bar{\mathcal{L}}_{-1}^k|h,h\rangle&=\frac{1}{2^k}\big(L_{1}(l)-lM_{1}(l)\big)^k|h,h\rangle=\big(L_{1}(l)\big)^k|h,h\rangle,
            \end{align}
since $L_{-1}(l)$ and $M_{-1}(l)$, $L_1(l)$ and $M_1(l)$ are pairwise commutative (see eq.\,\eqref{eq: BMS-like commutators before the flat limit}). To proceed, we calculate the following relation,
\begin{equation}
\mathcal{L}_{-1}^a L_{-1}^b(l)\overline{\mathcal{L}}_{-1}^c|h,h\rangle=-c(2h+c-1)\mathcal{L}_{-1}^{a-1}L_{-1}^b(l)\overline{\mathcal{L}}_{-1}^{c-1}|h,h\rangle+\mathcal{L}_{-1}^{a-1} L_{-1}^{b+1}(l)\overline{\mathcal{L}}_{-1}^c|h,h\rangle.
\end{equation}
where we have used the condition \eqref{eq: relation k} and the formulae (see eqs.\,\eqref{eq: L-1 and L1}, \eqref{eq: M-1 and L1})
\begin{align}
    \begin{aligned}
        \big[L_{-1}(l),L_{1}^c(l)\big] & =c(c-1)L_{1}^{c-1}(l)-2cL_{1}^{c-1}(l)L_0(l), \\
        \big[M_{-1}(l),L_{1}^c(l)\big] & =c(c-1)L_{1}^{c-2}(l)M_{1}(l)-2cL_{1}^{c-1}(l)M_0(l).
    \end{aligned}
\end{align}
Denoting $\ket*{a,b,c}:=\mathcal{L}_{-1}^a L_{-1}^b(l)\overline{\mathcal{L}}_{-1}^c|h,h\rangle$, we have the recursive relation
\begin{equation}
    \ket*{a,b,c}=-c(2h+c-1)\ket*{a-1,b,c-1}+\ket*{a-1,b+1,c}.
\end{equation}
It is worth noting that $s:=b-c$ is raised by $+1$ in both terms of the r.h.s., while $t:=b+c$ changes by $-1$ and $+1$ in each term.
Starting from
\begin{align}
    \ket*{k,0,k}=\mathcal{L}_{-1}^k\overline{\mathcal{L}}_{-1}^k|h,h\rangle\qquad(s=-k, t=k),
\end{align}
we must have the state $\ket*{0,a,a}=L_{-1}^a(l)L_1^a(l)|h,h\rangle$ ($a=0,1,\cdots,k$) after exactly $k$ successive recursions with $k-a$ times $\ket*{a,b,c}\to\ket*{a-1,b,c-1}$ type moves and $a$ times $\ket*{a,b,c}\to\ket*{a-1,b+1,c}$ type moves.
Along any path the total weight is
\begin{align}
    \prod_{i=1}^{k-a}\big(-(k-i+1)(2h+k-i)\big) & =(-1)^{k-a}\prod_{j=a+1}^{k}j(2h+j-1) \nonumber \\
    & = (-1)^{k-a}\frac{k!}{a!}\frac{\Gamma(k+2h)}{\Gamma(a+2h)}
\end{align}
where we relabelled $j=k-i+1$.
Crucially, this weight is independent of the order of two types of moves, so we finally obtain eqs.\,\eqref{eq: AdS descendants and fla descendants}\,\eqref{eq: relation des}.

\section{Another solution in the induced representation}
\label{append: another solution}
In this appendix, we present an alternative solution to the bulk equations for massive scalars \eqref{eq: origin equation}. However, we discover that the two-point function derived from this solution lacks short-distance divergence, making it an unsuitable candidate.

Consider again the equations \eqref{eq: origin equation} with the ansatz
\begin{equation}
    |\phi(0,0,0)\rangle=\sum_k q_k|k\rangle,
\end{equation}
where we have used the equation for $L_0$. Then the equation for $L_1$
\begin{equation}
    L_1|\phi\rangle=0,
\end{equation}
gives the recursive relation,
\begin{equation}
    q_{k-1}-k(k+1)q_k=0.
\end{equation}
With the initial condition $c_0=1$, the solution is
\begin{equation}
    c_k=\frac{1}{2(2)_{k-1}(3)_{k-1}}.
\end{equation}
Parallel to the discussion above, it is also the solution to
\begin{equation}
    L_{-1}|\phi\rangle=0,
\end{equation}
due to the fact that
\begin{equation}
L_{-1}^kL_{1}^k|\xi\rangle=L_{1}^kL_{-1}^k|\xi\rangle.
\end{equation}
The general solution at $(t,0,0)$ can be derived with further calculation
\begin{equation}
|\phi(t,0,0)\rangle=\sum_k\frac{(-2)^{-k} e^{i \xi  t} (i \xi  t)^{-k} \,
   _2\tilde{F}_2(1,1;2,1-k;-2 i t \xi )}{\Gamma (k+1)^2}|k\rangle.
\end{equation}
The bulk two-point function calculated from the inner product of this solution with the dual state $\ ^\vee\!\langle 0|$ is
\begin{equation}
  \ ^\vee\!\langle 0|\phi(t,0,0)\rangle=\frac{ie^{-it\xi}}{2t\xi}-\frac{ie^{it\xi}}{2t\xi},
\end{equation}
with the short distance behavior
\begin{equation}
    G(0,0,0)=1.
\end{equation}
So it is not the solution we want to reproduce the bulk two-point function.

\section{A toy model on the dual basis}
\label{app: toy model}
In this appendix, we want to give a toy model to illustrate the point that the bra state is defined with the usual conjugate and can be further organized in the dual basis. For the vectors in the complex vector space, the inner product is defined by the Hermitian form $\eta$
\begin{equation}
    (v,w):=\bar{v}^T\eta\  w.
\end{equation}
The adjoint $A^\dagger$ of the operator $A$ is
\begin{equation}
    (v,A w)=(A^\dagger v,w),
\end{equation}
for any $v,w$. Equivalently, in the matrix representation
\begin{equation}
    \bar{A}^T\eta=\eta A.
\end{equation}
Consider a simple non-diagonal $\eta$ as
\begin{equation}
    \eta=\begin{pmatrix} 0&1\\ 1&0 \end{pmatrix},
\end{equation}
and two operators
\begin{equation}
    \gamma=\begin{pmatrix} 0&0\\ 1&0 \end{pmatrix},\ \ \  \beta=\begin{pmatrix} 0&1\\ 0&0 \end{pmatrix},
\end{equation}
which are self-adjoint $\gamma^\dagger=\gamma,\beta^\dagger=\beta$ and satisfy the anti-commutation relation
\begin{equation}
    \{\beta,\gamma\}=1.
\end{equation}
Consider the basis
\begin{equation}\label{eq: toy basis}
    |0\rangle=(1\ \  0)^T,\ \ |1\rangle=(0\ \ 1)^T,
\end{equation}
and the bra states
\begin{equation}
    \langle0|=(1\ \  0),\ \ \langle1|=(0\ \ 1).
\end{equation}
We can write the action of $\gamma,\beta$ explicitly,
\begin{equation}
    \gamma|0\rangle=|1\rangle,\ \ \beta|0\rangle=0,\ \ \gamma|1\rangle=0,\ \ \beta|1\rangle=|0\rangle,
\end{equation}
and on the bra states
\begin{equation}
    \langle0|\gamma=\langle 1|,\ \ \langle 0|\beta=0,\ \ \langle 1|\gamma=0,\ \ \langle1|\beta=\langle 0|.
\end{equation}
With eq. \eqref{eq: dual basis1}, we find
\begin{equation}
      \ ^\vee\!\langle0|=\langle 1|,\ \   \ ^\vee\!\langle1|=\langle 0|,
\end{equation}
which are orthogonal to the basis \eqref{eq: toy basis},
\begin{equation}
      \ ^\vee\!\langle0|0\rangle=1,\   \ ^\vee\!\langle1|1\rangle=1,\   \ ^\vee\!\langle0|1\rangle=0,\   \ ^\vee\!\langle1|0\rangle=0.
\end{equation}
We then find the action of the $\gamma,\beta$ on the bra states in the dual basis,
\begin{equation}
    \ ^\vee\!\langle0|\gamma=0,\ \ \ ^\vee\!\langle0|\beta=\ ^\vee\!\langle1|,\ \ \ ^\vee\!\langle1|\gamma=0,\ \ \ ^\vee\!\langle1|\beta=\ ^\vee\!\langle0|.
\end{equation}
So we demonstrate that the dual basis corresponds to the relabeled bra state through the conventional conjugate relationship. The operators' actions adjust accordingly when applied to the dual basis. In practical applications, the connection between the dual basis and the standard basis in bra states can become quite intricate. However, identifying a bra state basis that is orthogonal to the desired ket state basis offers a promising candidate. This allows us to effectively represent any bra state.

    \bibliographystyle{JHEP}
    \bibliography{main}
\end{document}